\documentclass[11pt,aps,pre,notitlepage,longbibliography,tightenlines]{revtex4-1}
\usepackage{graphicx}
\usepackage{rotating}
\usepackage{bm}
\usepackage{amsmath}
\usepackage{amssymb}

\def\kon{k_{\rm on}}
\def\konc{k_{\rm on}c}
\def\koff{k_{\rm off}}
\def\koffe{k^{\rm end}_{\rm off}}

\def\rhom{\rho_{\rm max}}

\def\mum{$\mu$m}

\def\permin{ min$^{-1}$}
\def\pernm{ nM$^{-1}$}
\def\permum{$ \mu$m$^{-1}$}
\def\min1{$^{-1}$}

\begin{document} 
\title{Biophysics of filament length regulation by molecular motors}

\author{Hui-Shun Kuan} \affiliation{Program in Chemical Physics and
  Biofrontiers Institute, University of Colorado at Boulder}

\author{M. D.  Betterton} \affiliation{Department of Physics and
  Biofrontiers Institute, University of Colorado at Boulder}
\email{mdb@colorado.edu}

\date{\today}

\begin{abstract}
  Regulating physical size is an essential problem that biological
  organisms must solve from the subcellular to the organismal scales,
  but it is not well understood what physical principles and
  mechanisms organisms use to sense and regulate their size.  Any
  biophysical size-regulation scheme operates in a noisy environment2
  and must be robust to other cellular dynamics and fluctuations. This
  work develops theory of filament length regulation inspired by
  recent experiments on kinesin-8 motor proteins, which move with
  directional bias on microtubule filaments and alter microtubule
  dynamics.  Purified kinesin-8 motors can depolymerize
  chemically-stabilized microtubules. In the length-dependent
  depolymerization model, the rate of depolymerization tends to
  increase with filament length, because long filaments accumulate
  more motors at their tips and therefore shorten more quickly. When
  balanced with a constant filament growth rate, this mechanism can
  lead to a fixed polymer length.  However, the mechanism by which
  kinesin-8 motors affect the length of dynamic microtubules in cells
  is less clear.  We study the more biologically realistic problem of
  microtubule dynamic instability modulated by a motor-dependent
  increase in the filament catastrophe frequency. This leads to a
  significant decrease in the mean filament length and a narrowing of
  the filament length distribution. The results improve our
  understanding of the biophysics of length regulation in cells.
\end{abstract}
\maketitle

\section{Introduction}

A fundamental question in biology is how organisms control the size of
subcellular structures, cells, organs, and whole organisms.  The
physical principles underlying the sensing and control of size in
biology are not well understood; indeed, whether there are general
principles or mechanisms in size control is unclear
\cite{day00,hafen03,cook07}. In particular, the regulation of polymer
length is important for the organization of the cellular cytoskeleton.
Regulation of cytoskeletal filaments affects both the size of
subcellular organelles such as the mitotic spindle
\cite{goshima05,walczak96} and the structure of cells themselves
\cite{rivero96,revenu04}.  Microtubules are an important cytoskeletal
filament that contribute to cell structure, affect the distribution of
other cytoskeletal filaments, move chromosomes during cell division,
and serve as tracks for transport within the cell. This paper focuses
on the regulation of microtubule length.

Microtubules undergo complex, nonequilibrium polymerization dynamics,
known as \textit{dynamic instability}, characterized by stochastic
switching between distinct growing and shrinking states. When dynamic
microtubules polymerize \textit{in vitro} from purified tubulin
protein, dynamic instability leads to a broad distribution of polymer
lengths \cite{dogterom93}. However, in cells other proteins are also
present which modify microtubule dynamics, particularly at the plus
ends of microtubules \cite{akhmanova05}. This allows cells to control
tubulin polymerization dynamics to give microtubules of regulated
length. A relatively well-studied example of length regulation of
microtubule-based structures is the control of flagellar length in
\textit{Chlamydomonas reinhardtii}, where the assembly and disassembly
of the flagellum is controlled to give a fixed flagellar length
\cite{marshall05}. However, in general it is not well understood how
cells control the length of their microtubules.

Recently an example of physically-based microtubule length detection
and control was proposed, based on motor proteins that walk with
directional bias on a microtubule and shorten a stabilized
(non-polymerizing) microtubule from its plus end
\cite{gupta06,varga06,varga09}. If the motors are processive
(unbinding slowly from the microtubule), then a long filament can
accumulate large numbers of motors at its end, and the shortening rate
is high; for a short filament fewer motors reach the end, and
shortening slows.  This \textit{length-dependent depolymerization} has
been demonstrated for the kinesin-8 protein Kip3p moving on stabilized
microtubules.  The physical interactions of motors moving on the
filament allow a physical process (the rate of depolymerization) to
vary with the filament length, thereby allowing sensing of the length
\cite{varga06,varga09,hough09}. By coupling this length-sensitive
depolymerization with other processes (for example, a constant
filament growth rate) a specific filament length or narrow
filament-length distribution could in principle be achieved
\cite{varga06}.

To understand the biological relevance of length-dependent
depolymerization, it is important to make a connection between the
biophysically measured effects of purified proteins on stabilized
microtubules and the more complex situation in cells.  Stabilized
microtubules have little or no intrinsic length dynamics, while in
cells microtubules undergo dynamic instability.  Other proteins can
also modify microtubule dynamics.  Therefore, the kinesin-8
length-dependent depolymerization process will be affected by
microtubule length fluctuations and the presence of other proteins at
microtubule tips.  In general, any biophysical mechanism of length
regulation must be robust to noise in the cellular environment.

Recent work suggests that direct length-dependent depolymerization may
not be occurring in cells; instead, kinesin-8 proteins may act to
promote catastrophe (the transition from growing to shrinking).
Extensive experiments have demonstrated that deletions or knockdowns
of kinesin-8 proteins in cells result in longer microtubules and
mitotic spindles as well as an increase in chromosome loss in mitosis
\cite{west01,west02,garcia02a,garcia02b,savoian04,gupta06,mayr07,jaqaman10}.
Other work has shown that kinesin-8 activity is associated with
destabilization of microtubules and other alterations in microtubule
dynamics
\cite{west01,gandhi04,gatt05,gupta06,varga06,mayr07,unsworth08,tischer09,varga09,du10,peters10,wang10,gardner11a,masuda11,mayr11,stumpff11,su11,weaver11,erent12}.
While the molecular mechanisms of kinesin-8 protein function are not
clear, it appears that not all kinesin-8 proteins are able to
depolymerize stabilized microtubules \cite{grissom09,du10,erent12}.
Both experimental evidence \cite{gupta06,tischer09,gardner11a,erent12}
and modeling work \cite{brun09} suggest that in cells kinesin-8
proteins may act to promote catastrophe of dynamic microtubules.
Therefore, it is necessary to understand the consequences of
length-dependent changes in microtubule dynamic instability to predict
the effects of these motors in cells. This will improve our general
question of how length-sensing mechanisms are altered by fluctuations
and dynamics in biological systems.

Previous theory and modeling work has addressed aspects of kinesin 8
behavior and length regulation. Several papers have focused on
modeling the physical effects important to describe the
length-dependent depolymerization of otherwise static filaments
\cite{hough09,varga09,reese11} or filaments with simplified
polymerization kinetics \cite{govindan08,johann12,melbinger12}. To our
knowledge, previous work has not examined the effect of
catastrophe-promoting motors on the length distribution of
microtubules undergoing dynamic instability.  Tischer et al.~used a
similar formalism to that in this paper in a model for how
length-dependent microtubule catastrophe and rescue rates affect the
density of cargo-carrying motors along microtubules, an effect that
could be used to target cargo delivery to specific cellular locations
\cite{tischer10}.

In this paper, we develop a simplified physical theory to compare two
scenarios for length regulation: for \textit{length regulation by
  depolymerization} we calculate the steady-state length that is
reached by a constantly growing filament balanced by depolymerizing
motors, while for \textit{length regulation by altering catastrophe}
we consider filaments undergoing dynamic instability with alterations
in the dynamics due to motors. We consider two possible mechanisms of
motor action at the microtubule tip, both the minimal model in which
motor effects (depolymerization or catastrophe) increase in proportion
to the motor density \cite{hough09,reese11,melbinger12} and the
cooperative model in which motor effects (depolymerization or
catastrophe) increase in proportion to the flux of motors to the
filament end \cite{varga09}. These two models show important
differences in their effects on length regulation, suggesting that
cellular length regulation could be sensitive to the precise
mechanism. We find consistent qualitative agreement between mean-field
theory and stochastic simulation; in some parameter regimes the two
approaches agree quantitatively.

\section{Motor dynamics along filament}

\begin{figure}[t]
  \centering
  \includegraphics[width=16 cm]{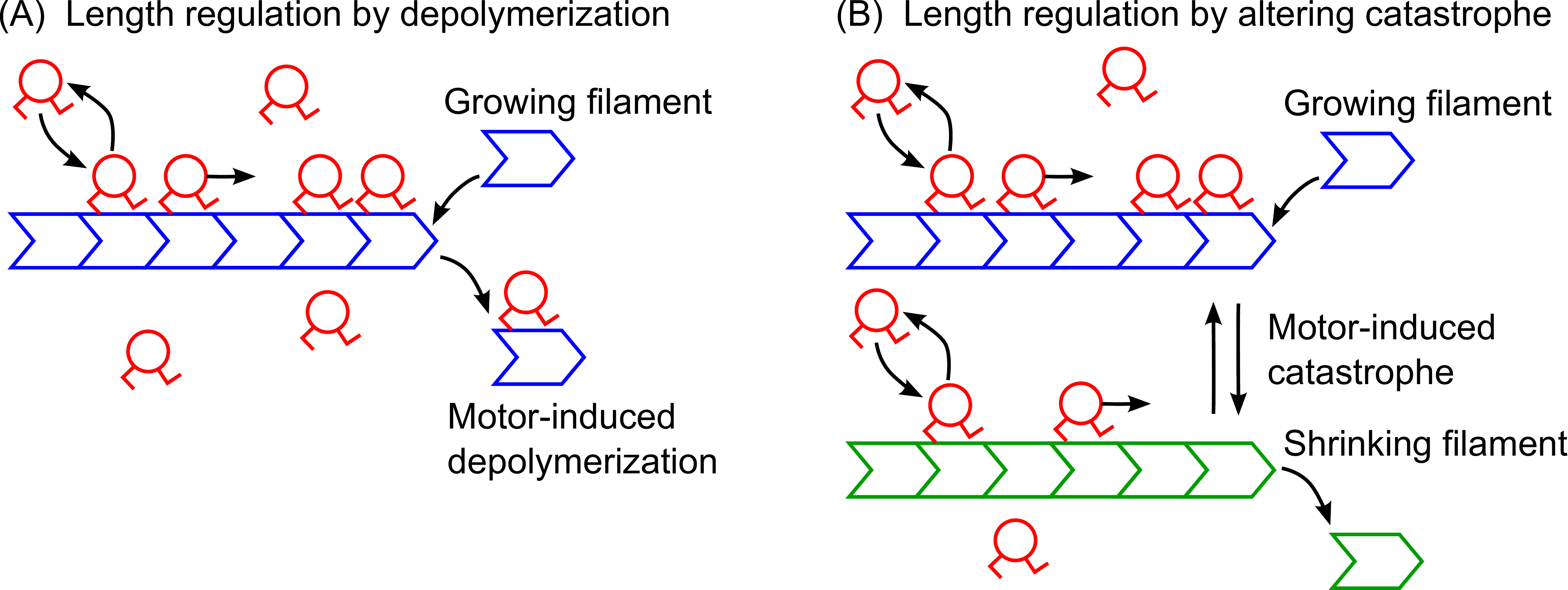}
  \caption{Schematic of the model. (A) Model of length regulation by
    depolymerization. Filament growth at the plus end is balanced by
    motor-induced depolymerization. Motors bind to and unbind from the
    filament, move toward the filament plus end, and catalyze removal
    of filament subunits from the plus end. This leads to a
    length-dependent depolymerization rate, so a single filament
    length is favored, depending on the model parameters.  (B) Model
    of length regulation by altering catastrophe.  The filament
    undergoes dynamic instability at its plus end, characterized by
    stochastic transitions between growing (blue) and shrinking
    (green) states. Motors bind to and unbind from the filament, move
    toward the filament plus end, and catalyze catastrophe (transition
    from the growing to the shrinking state) at the filament plus end.
    Motor effects make the catastrophe frequency length dependent,
    which leads to a broad distribution of filament lengths determined
    by the properties of the motors.  }
  \label{fig:model}
\end{figure}

The mean-field density of the motors along the filament, $\rho(x,t)$, in
units of motors per unit length is described by \cite{parmeggiani04}
\begin{equation}
  \frac{\partial \rho}{\partial t} = -v \frac{\partial }{\partial x}
  \left[\rho\left(1-\frac{\rho}{\rhom} \right) \right]  + \konc  \left(1-\frac{\rho}{\rhom} \right)-  \koff \rho.    
\label{eq:rhoxgen}
\end{equation}
On the right hand side, the first term describes biased motion of the
motors with speed $v$, where crowding effects reduce the motor flux
\cite{parmeggiani04} and $\rhom$ is the maximum possible motor
density. The second term describes binding of motors to unoccupied
sites at rate per unit length $\kon c $. The third term describes
unbinding of motors from occupied sites at rate $\koff$.  This
formulation assumes a continuum limit in which the lattice spacing $a
\to 0$ so that the motor density can be treated as continuous in $x$.
The bulk motor concentration $c$ is assumed constant, i.e., the
binding of motors to the filament is assumed not to deplete the pool
of motors in the bulk.  Note that we do not consider protofilament
interactions within a microtubule, so we are effectively considering a
single-protofilament microtubule (figure \ref{fig:model}).

For relatively low motor density, we neglect crowding effects in the
drift term, which makes the density equation linear. In addition, we
work with the motor fractional occupancy $p(x,t) = \rho(x,t) /\rhom$,
so the density equation can be written
\begin{equation}
  \frac{\partial p}{\partial t} = -v \frac{\partial p}{\partial
    x}  +  \frac{\konc}{\rhom}  \left(1-p \right)-  \koff p.    
\label{eq:rhoxsim1}
\end{equation}
With the initial condition $p(x,t)=0$ at time $t=0$ and the boundary
condition $p(x=0,t)=0$, the solution to this equation is
\begin{equation}
  \label{eq:rhosoln}
  p(x,t) =  \begin{cases}
 p_0(1-e^{-t/\tau}), &x \ge v t \mbox{      (short time)} \\
 p_0(1-e^{-x/\lambda}), &x < v t \mbox{      (long time)} \\
\end{cases}
\end{equation}
The steady-state occupancy away from the filament ends is $p_0 =
\konc/ (\koff \rhom + \konc)$, the time scale $\tau = 1/(\koff +
\konc/\rhom)$, and the length scale $\lambda = v/( \koff +
\konc/\rhom) = v \tau$. At long time, the density approaches the
steady state profile which is constant away from the filament end but
has a boundary layer for small $x$ where transport effects and
boundary conditions change the motor density away from $p_0$
\cite{parmeggiani04,nowak07,govindan08}.

\section{Length regulation by depolymerization}

Here we study the regulation of filament length assuming the motors
directly depolymerize the filament from its plus end, an effect which
is balanced by a constant rate of filament growth (figure
\ref{fig:model}A). This approach neglects fluctuations due to
microtubule dynamic instability, and so the resulting length is
deterministically reached.  We suppose that a filament grows with
constant speed $u$.

We consider two simple models for the dynamics of the plus end end.
(We assume that the filament minus end is not dynamic.) For
\textbf{density-controlled depolymerization}, the motor-induced
depolymerization rate is proportional to the motor occupancy at the
end \cite{hough09,reese11,melbinger12}. For \textbf{flux-controlled
  depolymerization}, the motor-induced depolymerization rate is
proportional to the motor flux to the end \cite{varga09,reese11}.  We
assume that the motors move faster than the growth ($v>u$), so the
motors track the end as observed experimentally. We therefore assume
that the motor occupancy away from the filament end reaches the
steady-state value $p(x) = p_0(1-e^{-x/\lambda})$.

Dimensional analysis suggests that the filament length reached should
be related the boundary layer length scale $\lambda = v/( \koff +
\konc/\rhom) = v \tau$, the obvious length scale that can be
constructed from the rates in the problem. However, the steady-state
filament length is quite different from $\lambda$ and is controlled by
the dynamics at the end of the filament.

\begin{figure}[t]
  \centering
  \includegraphics[width=5 cm]{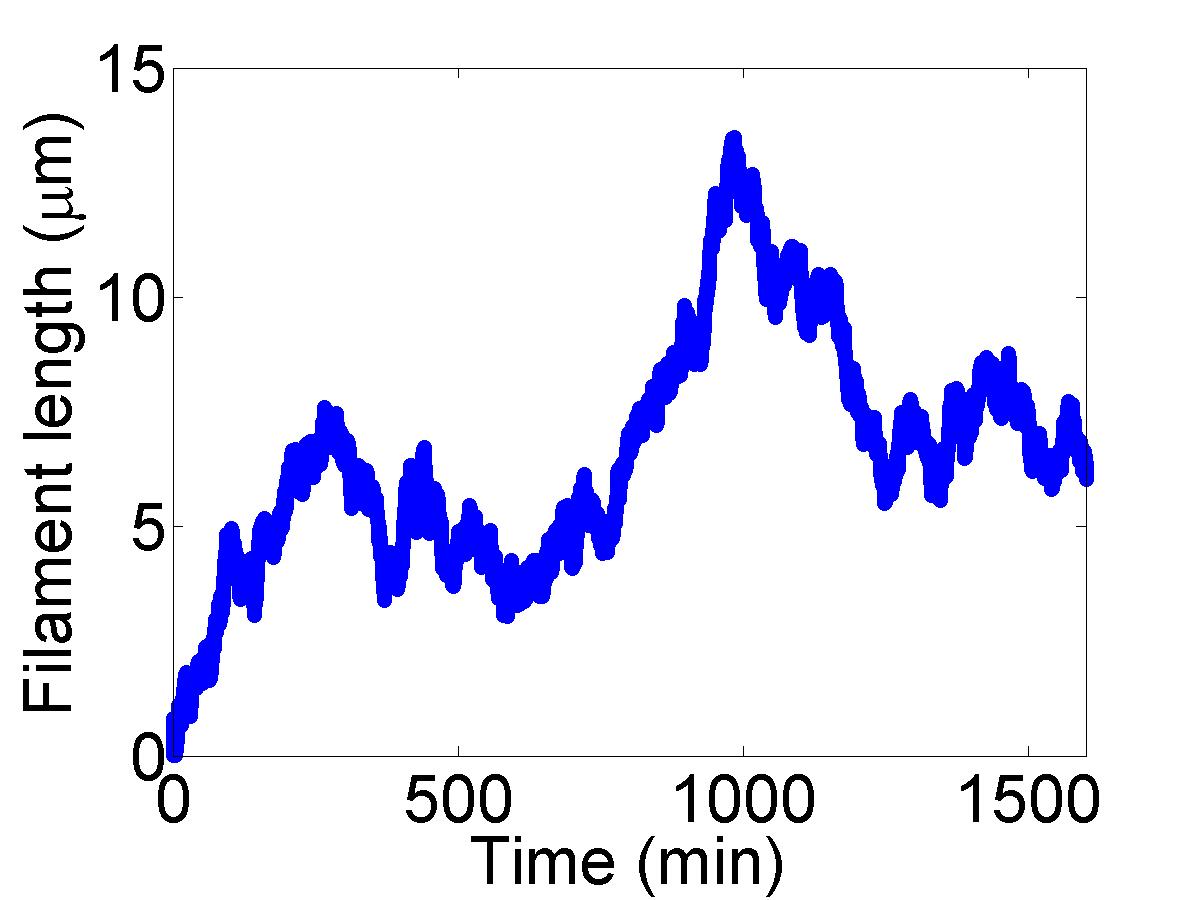}
  \includegraphics[width=5 cm]{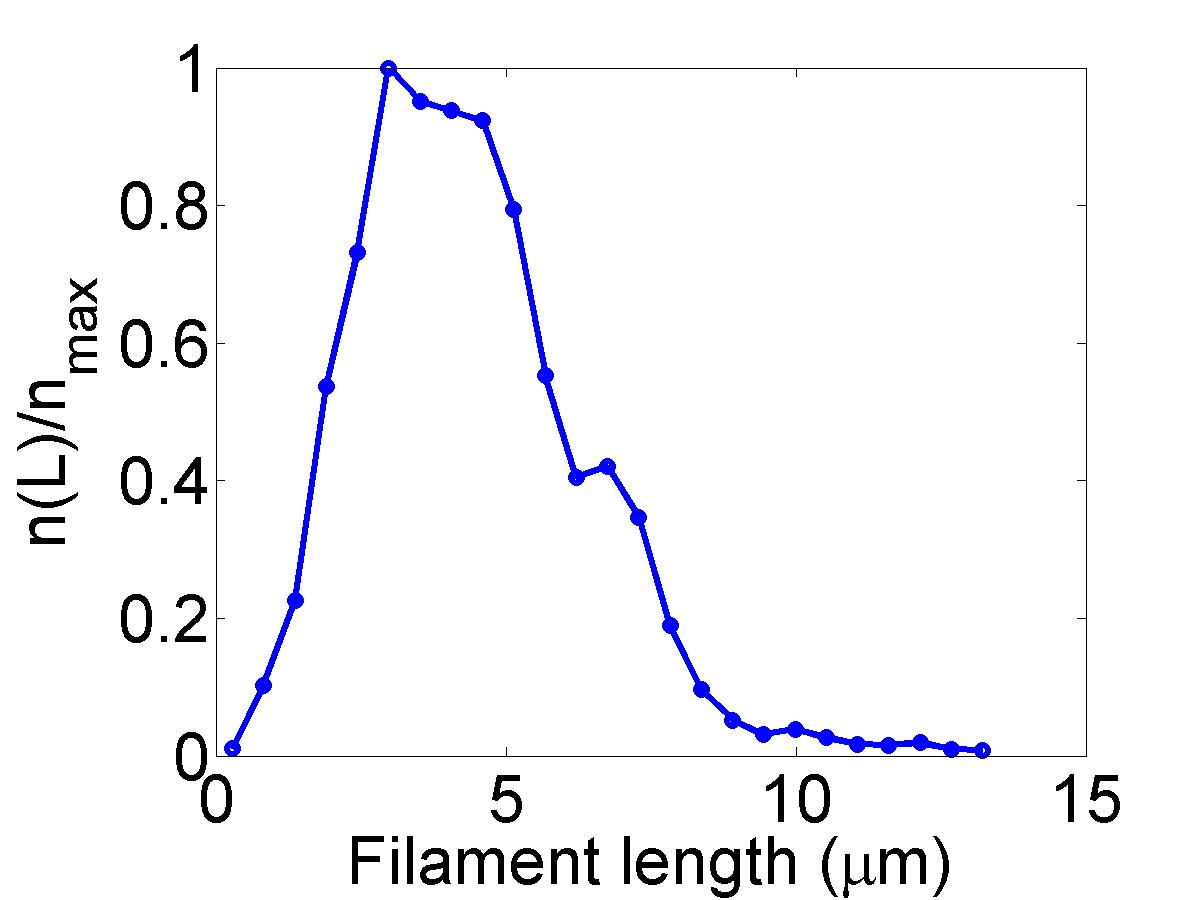}
  \includegraphics[width=5 cm]{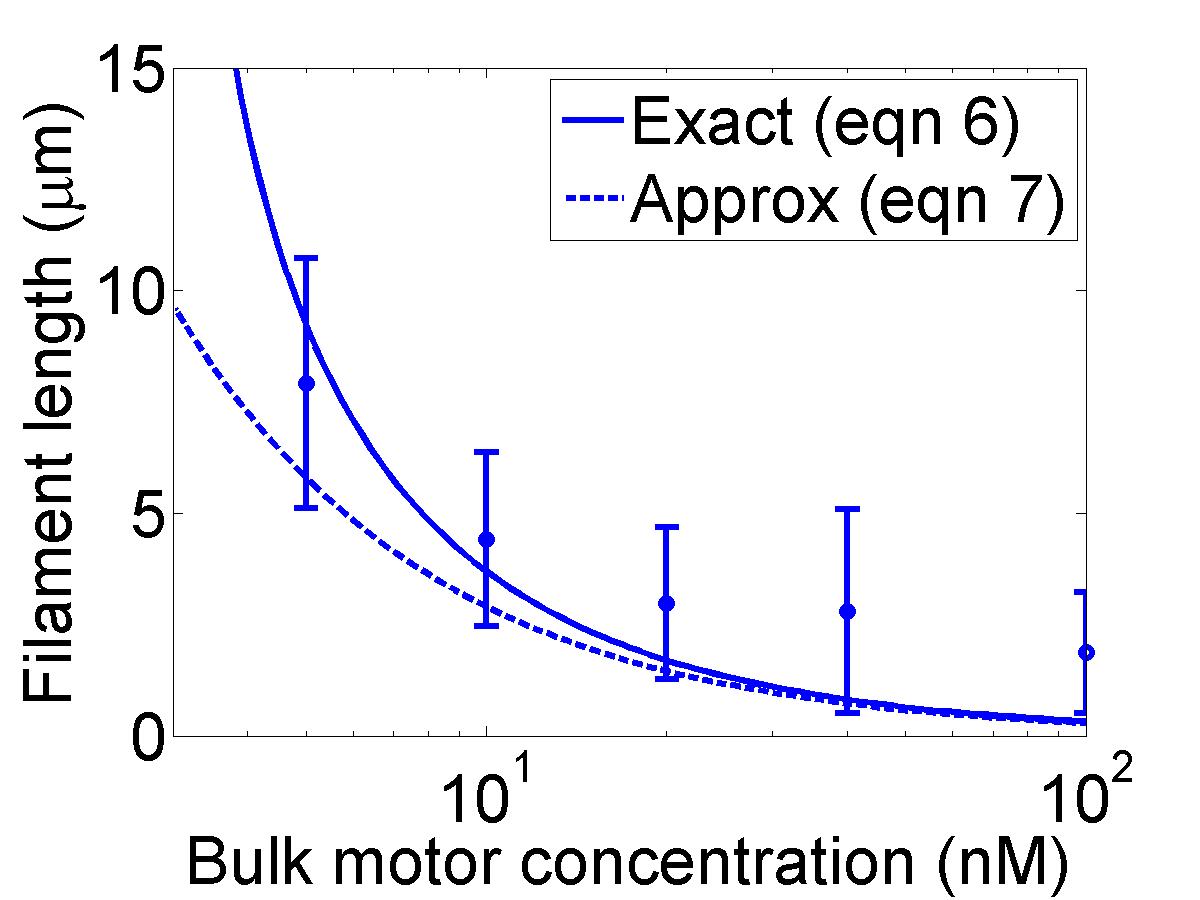}
  \includegraphics[width=5 cm]{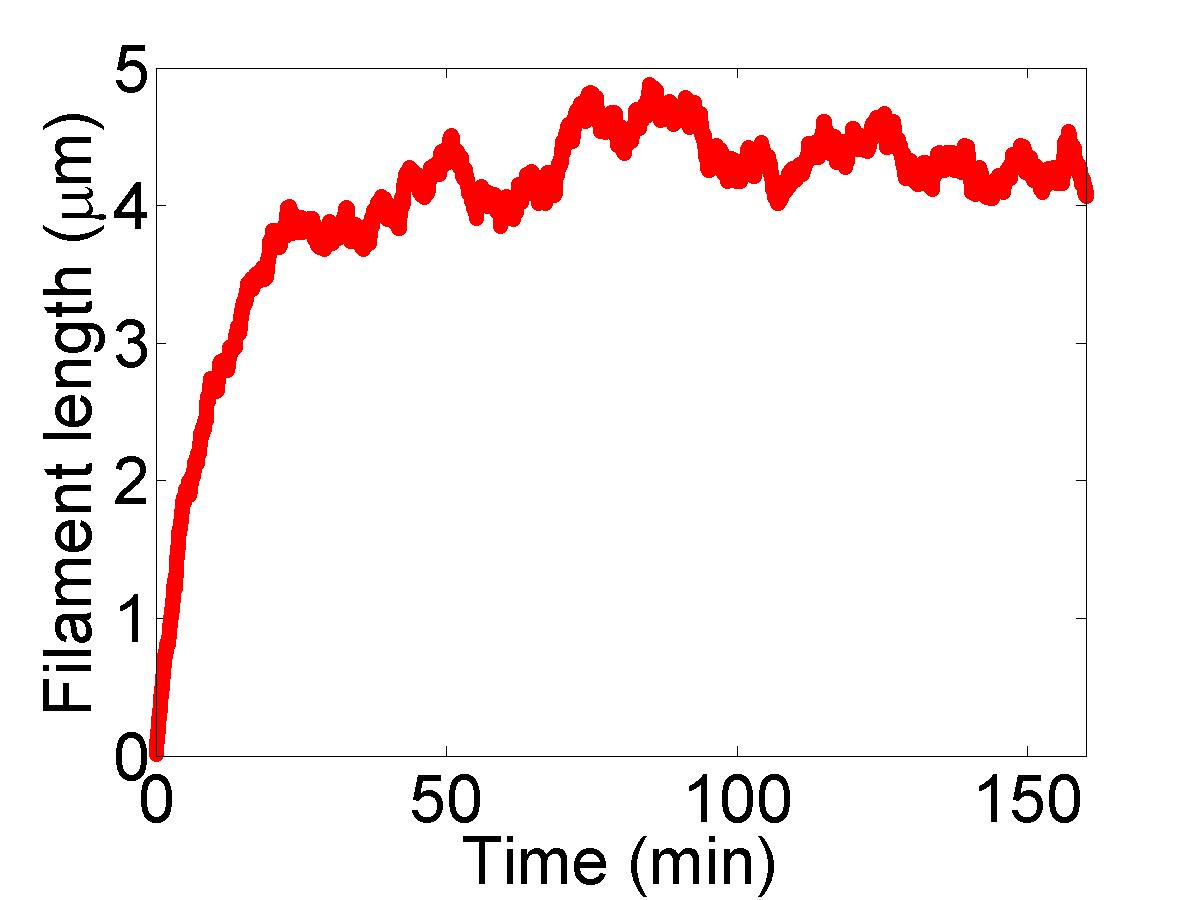}
  \includegraphics[width=5 cm]{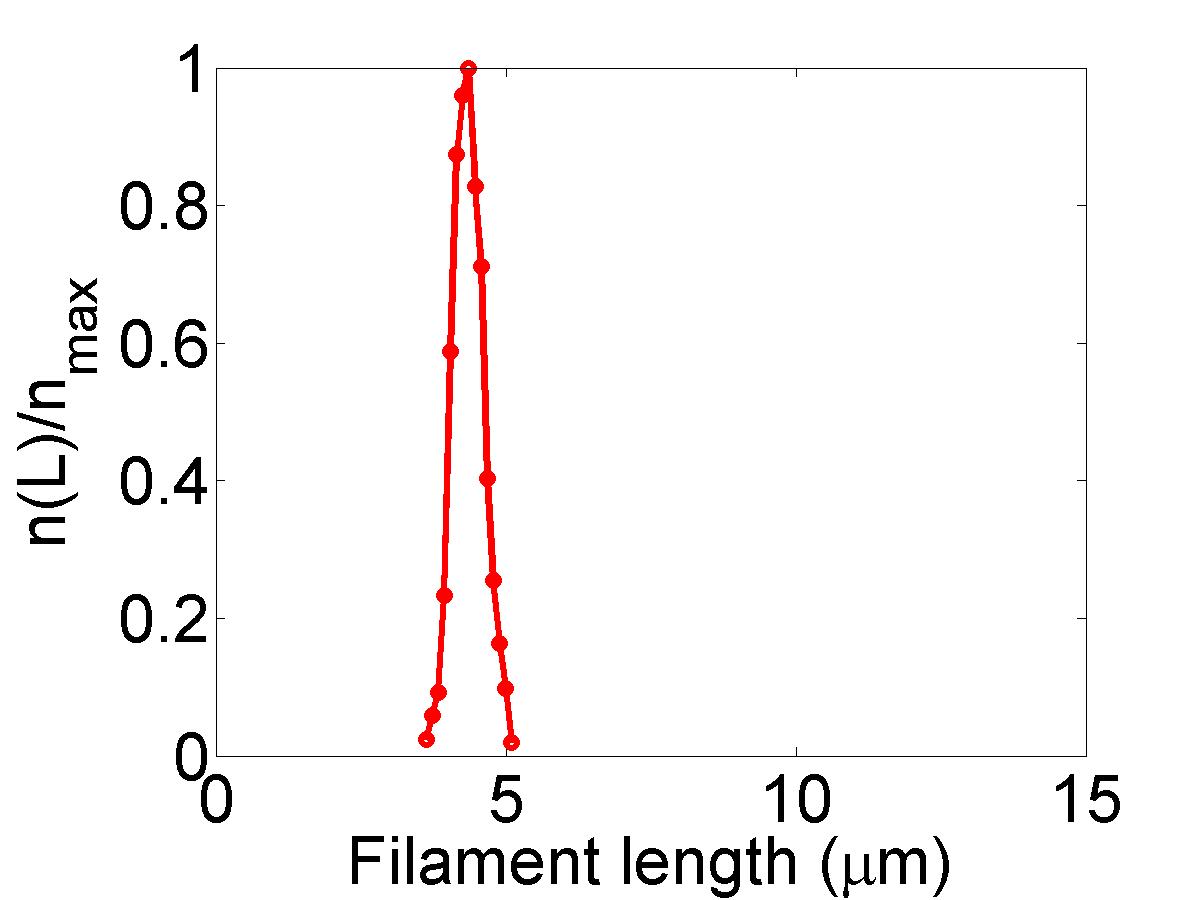}
  \includegraphics[width=5 cm]{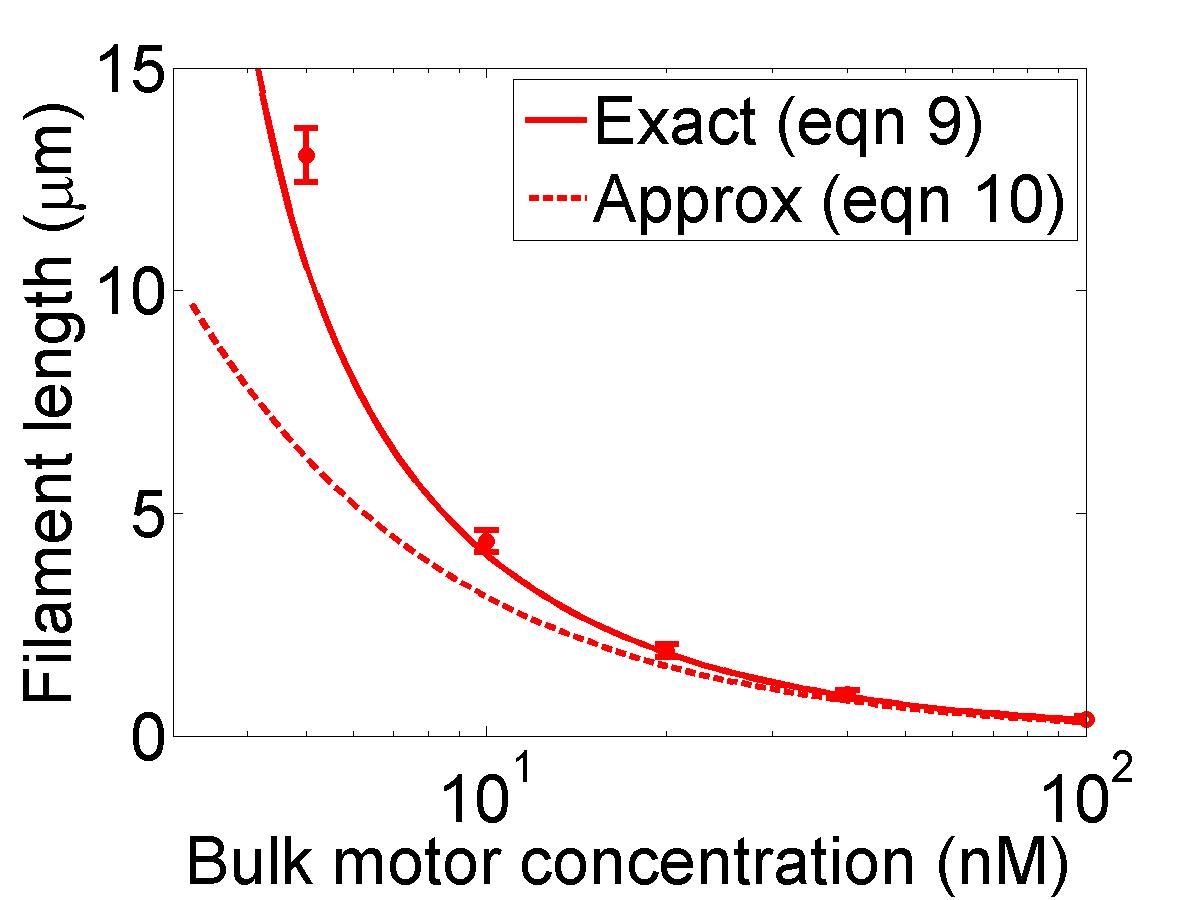}
  \caption{Filament dynamics and steady-state filament length for
    length regulation by depolymerization.  Blue (top),
    density-controlled depolymerization; red (bottom), flux-controlled
    depolymerization.  Left, example trace of filament length versus
    time in the stochastic simulation for $c=10$ nM. Middle,
    normalized filament length distribution in the stochastic
    simulation for $c=10$ nM, averaged from 10 stochastic simulations
    after removal of the initial transient. Right, steady-state
    filament length predictions of mean-field theory and the
    stochastic simulation (error bars are standard deviations of
    steady-state length distributions).  The steady-state length
    varies rapidly with the bulk motor concentration. The two models
    give similar predictions, and the approximate expressions (eqns 7
    and 10) for the steady-state length are within a factor of two of
    the exact expressions (eqns 6 and 9) except for very low bulk
    motor concentrations.  The mean-field theory uses the parameters
    $v = 3$ \mum \permin, $\kon = 2$ \pernm \permum \permin, $\koff =
    0.25$ \permin, $\koffe = 1.45$ \permin, $w = 1.025$ \mum \permin,
    $a = 8$ nm, $\delta = 8$ nm, and $\rhom = 125$ \permum; for the
    density-controlled model $u=1.0$ \mum \permin\ while for the
    flux-controlled model $u=0.5$ \mum \permin.  The stochastic
    simulation uses the same parameters except $w=1.5$ \mum \permin\
    and $\koffe = 1$ \permin\ for the density-controlled model.}
  \label{fig:lss}
\end{figure}

This model is related to recent work that also considered the balance
between depolymerizing motors and filament kinetics described by
constant growth \cite{govindan08,melbinger12} or treadmilling
\cite{johann12}. The model here is simplified compared to the previous
work, to allow the derivation of analytic expressions for the length
achieved and to allow comparison to the results for filaments
undergoing dynamic instability.

\subsection{Density-controlled depolymerization}

In the density-controlled model, we assume that the depolymerization
rate is proportional to the motor density at the terminal site of the
microtubule \cite{hough09}.  We assume that only the motor occupancy
at the last site of the filament is important for depolymerization,
i.e., we neglect possible cooperative effects. Define $p_e(t)$ to be
the average motor occupancy at the last site on the filament, and the
filament length is $L(t)$. The mean-field density-dependent
depolymerization model is:
\begin{eqnarray}
  \dot{ p_e} &=&  \left(\frac{v-\dot{ L}}{a}\right)p(L-a,t)(1-p_e)   
  - \koffe p_e,  \label{eq:dndt} \\
  \dot{L} &=&  u - w p_e. 
\label{eq:dldt}
\end{eqnarray}
The first term on the right side of equation \eqref{eq:dndt}
represents the stepping of motors from the site adjacent to the end to
the terminal site on the filament, at rate $(v-\dot{L})/a$. The
density at the penultimate site on the filament is $p(L-a,t)$, where
$a$ is the lattice spacing, assumed small.  If the motor dynamics at
the end are faster than the typical time scale of density changes in
the bulk of the filament, we can treat the motor density away from the
end quasi-statically assuming it is unaffected by the end dynamics.
Thus, we write $p(L-a,t)\approx p(L,t)$, where $p(x,t)$ is the motor
occupancy for a region far from the filament end.  Because the
kinetics of motor removal may be different at the end of the filament,
we include crowding effects at the last site even though they are
neglected elsewhere along the microtubule.  The second term on the
right side of equation \eqref{eq:dndt} describes unbinding of the
motor at the end.  In equation \eqref{eq:dldt}, the filament lengthens
at constant speed $u$ and shortens at a rate proportional to the motor
density at the end, with a maximum speed $w$.

Note that in this model the unbinding of the motor from end of the
filament (controlled by the term with rate $\koffe p_e$ in equation
\eqref{eq:dndt}) is decoupled from the depolymerization rate
(controlled by the term with rate $w p_e$ in equation
\eqref{eq:dldt}). This means that we allow processive
depolymerization, with a single motor able to remove an average of
$w/\koffe$ monomers from the filament.

The steady-state length in the density-controlled model $L_{\rm den}$
is reached when $\dot{L} = 0$ and $\dot{ p_e}=0$, which implies $p_e =
u/w$ and
\begin{eqnarray}
  \label{eq:lengthss}
  L_{\rm den} &=& -\lambda \ln \left[ 1- \frac{u a \koffe}{p_0 v
      (w-u)}\right] \\     
  &\approx & \frac{u }{\konc}  \left(\frac{\rhom a \koffe}{w-u} \right)
\label{eq:lengthdenapprox} 
\end{eqnarray}
The approximate solution in equation \eqref{eq:lengthdenapprox}
applies when the second term inside the logarithm of equation
\eqref{eq:lengthss} is small.  Note that there is no steady-state
solution if either $w < u$ (in this case the motors can never remove
dimers quickly enough to keep up with the growth) or $ u a \koffe/(p_0
v (w -u)) \ge 1$. Therefore the motor occupancy must be larger than
the critical value $p_{0c} = u a \koffe /(v (w-u))$ for a steady-state
length to occur, implying a minimum bulk motor concentration of $c_c =
\koff \rhom a \koffe/(\kon [w (w/u-1)-a \koffe]) $. In practice given
measured motor parameters, reasonable values of the steady-state
length require $w$ quite similar to $u$; the requirement for such
fine-tuning of the depolymerization rate suggests that this
length-regulation mechanism is not highly robust (figure
\ref{fig:lss}).

We show the dependence of the steady-state length on the bulk motor
concentration in figure \ref{fig:lss}. We use parameters similar to
those found in experiments \cite{varga09,gupta06}.  The results of
stochastic simulation of density-controlled depolymerization agree
qualitatively with the predictions of the mean-field theory (figure
\ref{fig:lss}, details of the simulations are described in section
\ref{sim}). The best agreement occurs when the stochastic simulation
uses a slightly larger motor-induced depolymerization speed $w$ and a
slightly lower motor unbinding rate from the filament end $\koffe$
than the mean-field theory. In this case the mean filament lengths
reached in the two models are similar, but the stochastic simulation
shows large fluctuations about the mean length.  This is intuitively
reasonable since in this model depolymerization is controlled by the
motor occupancy at the terminal site of the filament, which undergoes
significant fluctuations.

\subsection{Flux-controlled depolymerization}

Varga et al.~found that their experimental data for depolymerization
of stabilized microtubules by Kip3p are consistent with filament
depolymerization being determined by the flux of motors to the end
\cite{varga09}.  In this model, a motor would in principle remain
bound to the filament end forever, unless displaced from the tip by
the arrival of another motor. When unbinding, each motor is assumed to
shorten the microtubule by a length $\delta$ (where $\delta$ could
equal the lattice spacing $a$ if each motor removes exactly one
tubulin dimers, but could differ from $a$ depending on the motor
depolymerization processivity). Therefore the depolymerization speed
is $w= \delta J(L)$, where $J(L)= p(L) \rhom (v-\dot{L})$ is the flux
of motors to the end of the filament.  Note that steric interactions
between motors that decrease the flux are neglected here.  The length
of the microtubule changes in time according to
\begin{equation}
  \label{eq:fluxvarga}
  \dot{L}= u -w = u- \delta \rhom
  p(L)\left(v-\dot{L} \right) 
\end{equation}
At steady state, $\dot{L}=0$ and the motor occupancy is the
steady-state value. Therefore $u = \delta \rhom v p(L_{\rm ss})$, and
the steady-state length in the flux-controlled model is
\begin{eqnarray}
  L_{\rm flux} &=& -\lambda \ln \left[ 1-  \frac{u}{ p_0 v \delta  \rhom} 
  \right] \\ 
  &\approx& \frac{u}{\konc } \left( \frac{1}{\delta } \right). \label{eq:lengthfluxapprox} 
\end{eqnarray}
As above, the approximate solution applies when the second term inside
the logarithm is small. Note that there is no steady-state solution if
$ u /(v \delta p_0 \rho_m ) \ge 1$. This implies that the motor
density must be larger than the critical value $ p_{0c} = u/(v \delta
\rhom) $ for a steady-state length to occur; this corresponds to a
critical bulk motor density $c_c = \koff \rhom u/(\kon (v \delta \rhom
- u))$.  This requires $v \delta \rhom > u$; in practice, for
parameters for the budding-yeast motor Kip3 $v \delta \rhom$ must be a
few times $u$ to get steady-state lengths of a few microns.

We show the dependence of the steady-state length on the bulk motor
concentration in figure \ref{fig:lss}.  The results of stochastic
simulation of flux-controlled depolymerization agree quantitatively
with the predictions of the mean-field theory for identical parameters
(figure \ref{fig:lss}, details of the simulations are described in
section \ref{sim}). Compared to the density-controlled
depolymerization model, the flux-controlled depolymerization model
shows decreased fluctuations and a relatively narrow length
distribution. This may occur because in this model depolymerization is
controlled by the motor flux to end of the filament, which is a
collective property of multiple motors.

The structures of the steady-state solutions are similar in the two
models, having the approximate form (equations
\eqref{eq:lengthdenapprox} and \eqref{eq:lengthfluxapprox}) $L \sim
u/\konc$ times a factor with units of inverse length related to how
motors are removed from the end. These approximations to $L$ make
clear the strong dependence of the steady-state length on the bulk
motor concentration, implying that length regulation by this mechanism
requires tight regulation of the motor concentration $c$. In the
density-controlled model the motor unbinding rate from the end of the
filament and the difference $w-u$ between the maximum speed of
depolymerization and the filament growth rate are important in
controlling the length reached. In the flux-controlled model the
steady-state length takes a simple form, depending primarily on $u$,
$\konc$, and $\delta$.

In both cases, there is a minimum motor occupancy required to reach a
steady-state length, as expected, because a minimum number of motors
is required for depolymerization to balance polymerization. The
steady-state filament length is quite different from the dimensional
length scale $\lambda$ which characterizes the motor density profile.

\section{Length regulation by altering catastrophe}

In cells, microtubule filaments typically don't grow constantly as in
the simple model above, but instead undergo dynamic instability,
characterized by long-lived growing and shrinking regimes with
transitions between these two states.  Studies of Kip3p in cells
\cite{gupta06} and \textit{in vitro} \cite{gardner11a} and of other
kinesin-8 motors \cite{tischer09,erent12}, as well as modeling work
\cite{brun09} suggest that in cells these proteins may act to promote
catastrophe (the transition from growing to shrinking) of dynamic
microtubules. Therefore, it is necessary to understand the
consequences of length-dependent changes in filament dynamics (rather
than merely shortening) to predict the effects of these motors in
cells.

Here we develop a theory of motors that promote filament catastrophe
in a length-dependent manner.  The number probability density $n(L) =
n_G(L) + n_S(L)$ for filaments of length $L$ is made up of two
populations, growing (G) and shrinking (S) filaments. The total number
of filaments is $N = \int n(L) \ dL$.  In this model, we neglect
pauses (neither growth nor shrinkage) exhibited by dynamic
microtubules in cells. The distributions evolve according to
\begin{eqnarray}
  \frac{\partial n_G}{\partial t} &=& -u \frac{\partial n_G}{\partial L} -
  f_c n_G + f_r n_S\\
  \frac{\partial n_S}{\partial t} &=& w \frac{\partial n_S}{\partial
    L} + f_c n_G -
   f_r n_S \label{eq:mtdistshrink}
\end{eqnarray}
The terms in the first equation represent filament growth with speed
$u$, catastrophe with frequency $f_c$, and rescue with frequency
$f_r$. The terms in the second equation represent filament shortening
with speed $w$, catastrophe, and rescue. At steady state, $ u
\frac{\partial n_G}{\partial L} = w \frac{\partial n_S}{\partial L}$.
The solution to this equation (consistent with the boundary condition
that the number of filaments drops to zero as $L \to \infty$) is $n_S
= (u/w) n_G$.  Then the steady-state equation for the total number of
filaments $n = n_G + n_S$ simplifies to
\begin{equation}
  \label{eq:ngss}
   \frac{\partial n}{\partial L} =  -\left( \frac{f_c}{u}-
     \frac{f_r }{w}\right) n
\end{equation}
If the catastrophe and rescue rates are spatially constant, the
microtubule length distribution is exponential, $ n(L) = n_{0}
\exp(-(f_c/u-f_r/w)L)$, so the distribution is a bounded exponential
if $f_c/u>f_r/w$ and has characteristic length $u w/(f_c w - f_r u)$.

In the case of  length-dependent rates, we have the formal solution
\begin{equation}
  \label{eq:ldistspat}
    \ln n = - \int dL\ \left( \frac{f_c}{u}-
     \frac{f_r }{w}\right)
\end{equation}
Here, we assume that only the catastrophe rate $f_c(L)$ varies with
length (as observed for the kinesin-8 motors Klp5/6 in fission yeast
cells \cite{tischer09}), and other rates are all constant. Then
\begin{equation}
  n= n_{0} e^{f_r L/w } \exp\left(- \frac{1}{u} \int dL\ f_c(L) \right)
\end{equation}

\begin{figure}[t]
  \centering
 \includegraphics[width=5 cm]{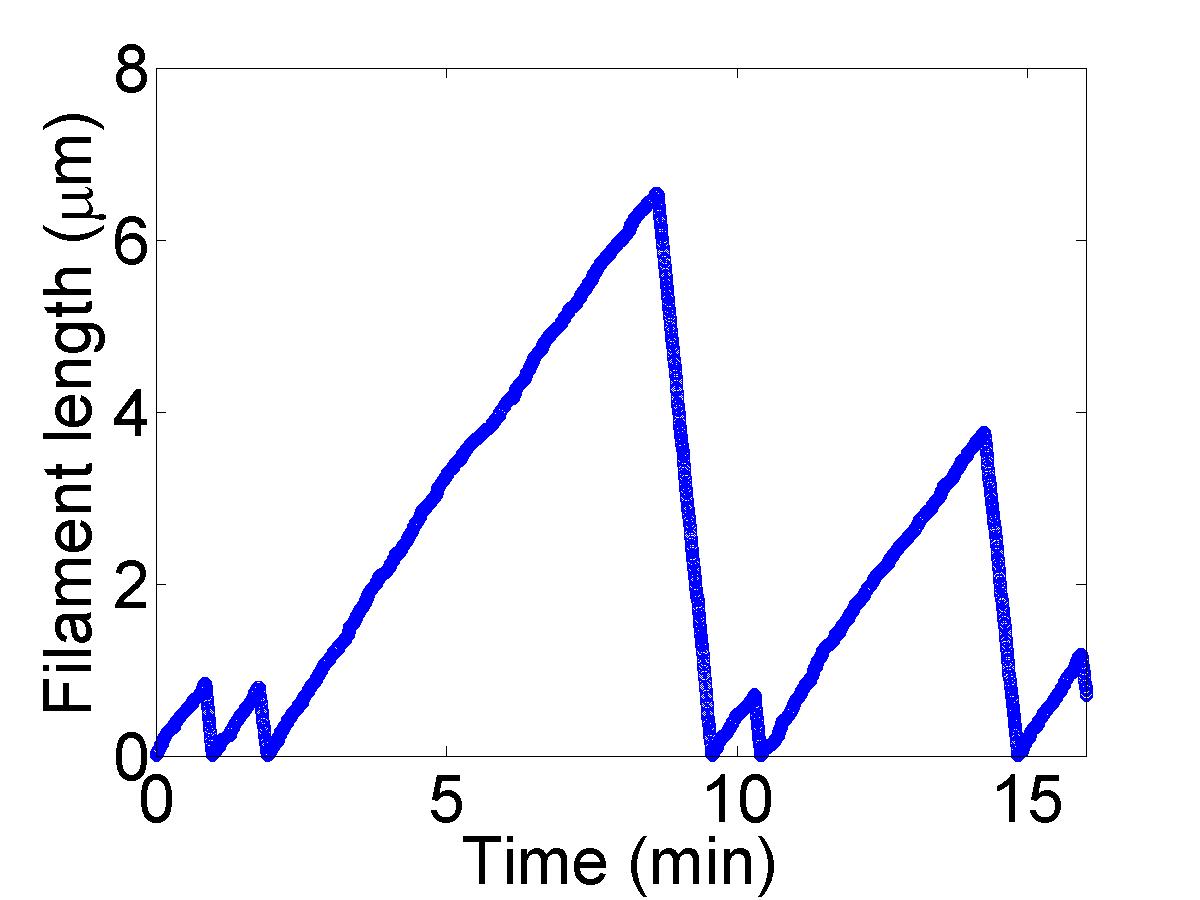}
 \includegraphics[width=5 cm]{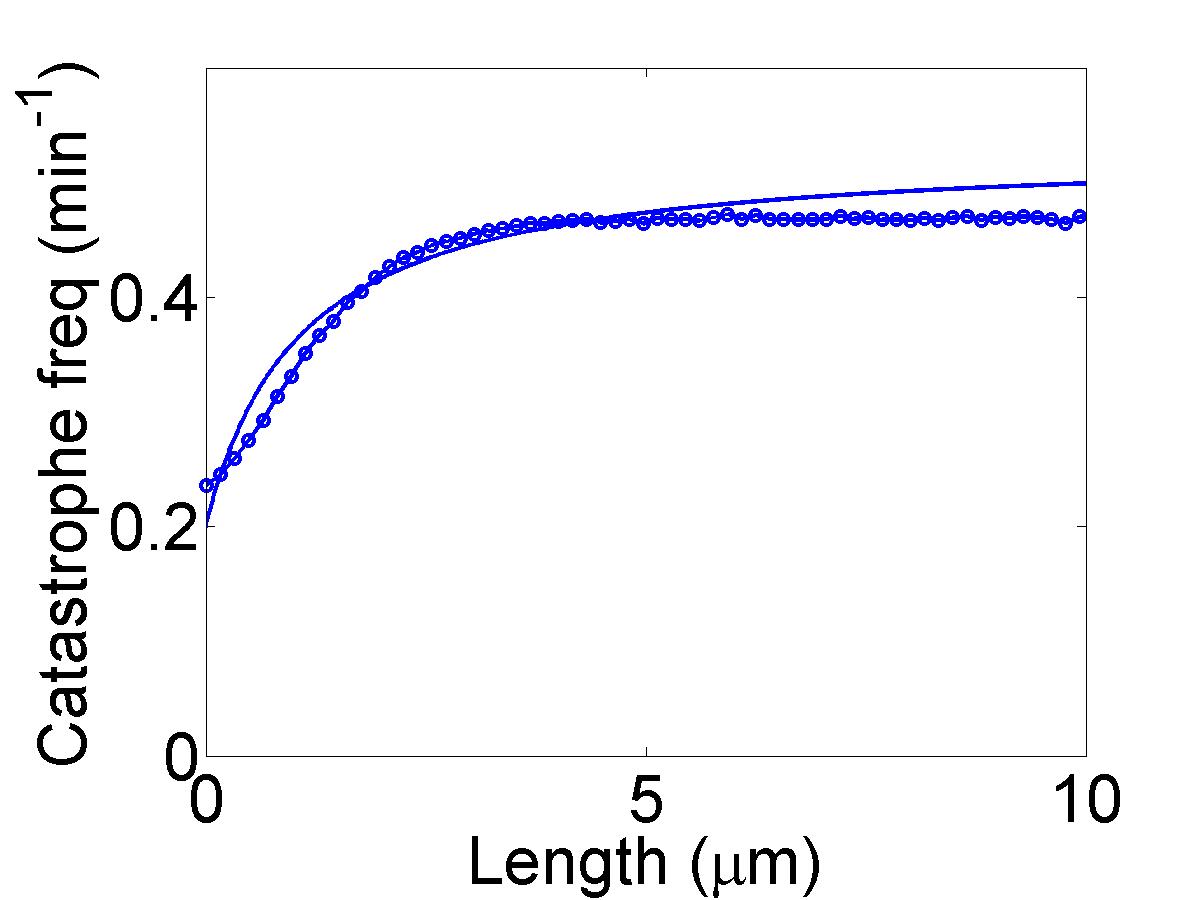}
 \includegraphics[width=5 cm]{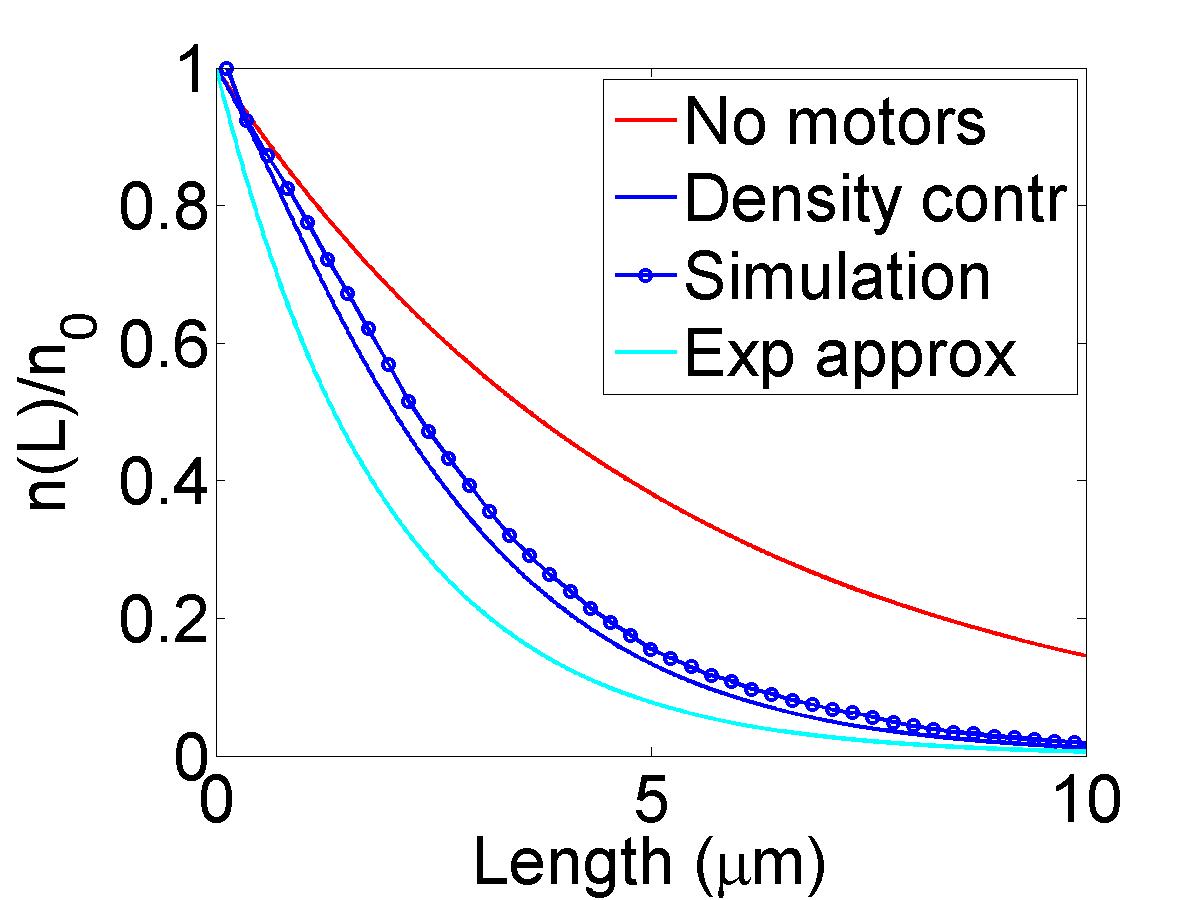}
 \includegraphics[width=5 cm]{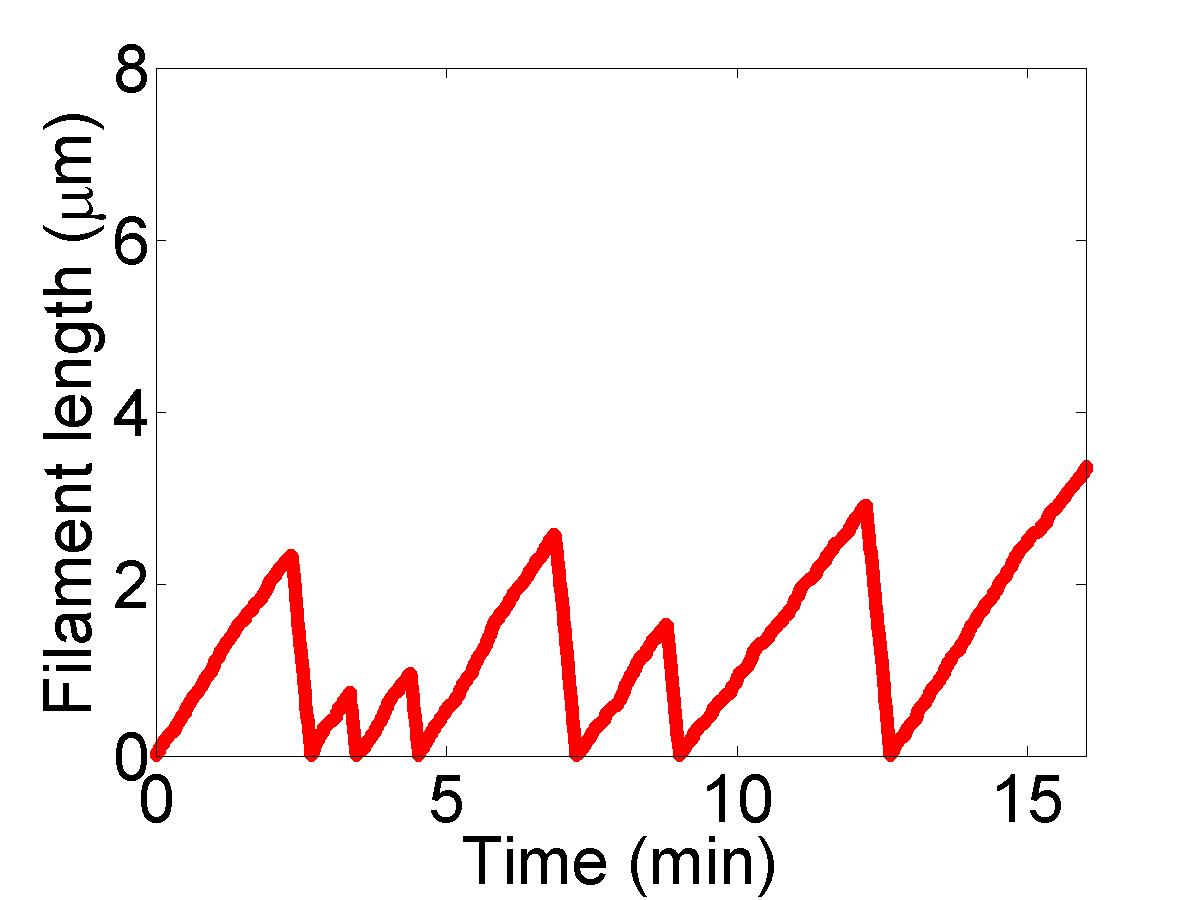}
 \includegraphics[width=5 cm]{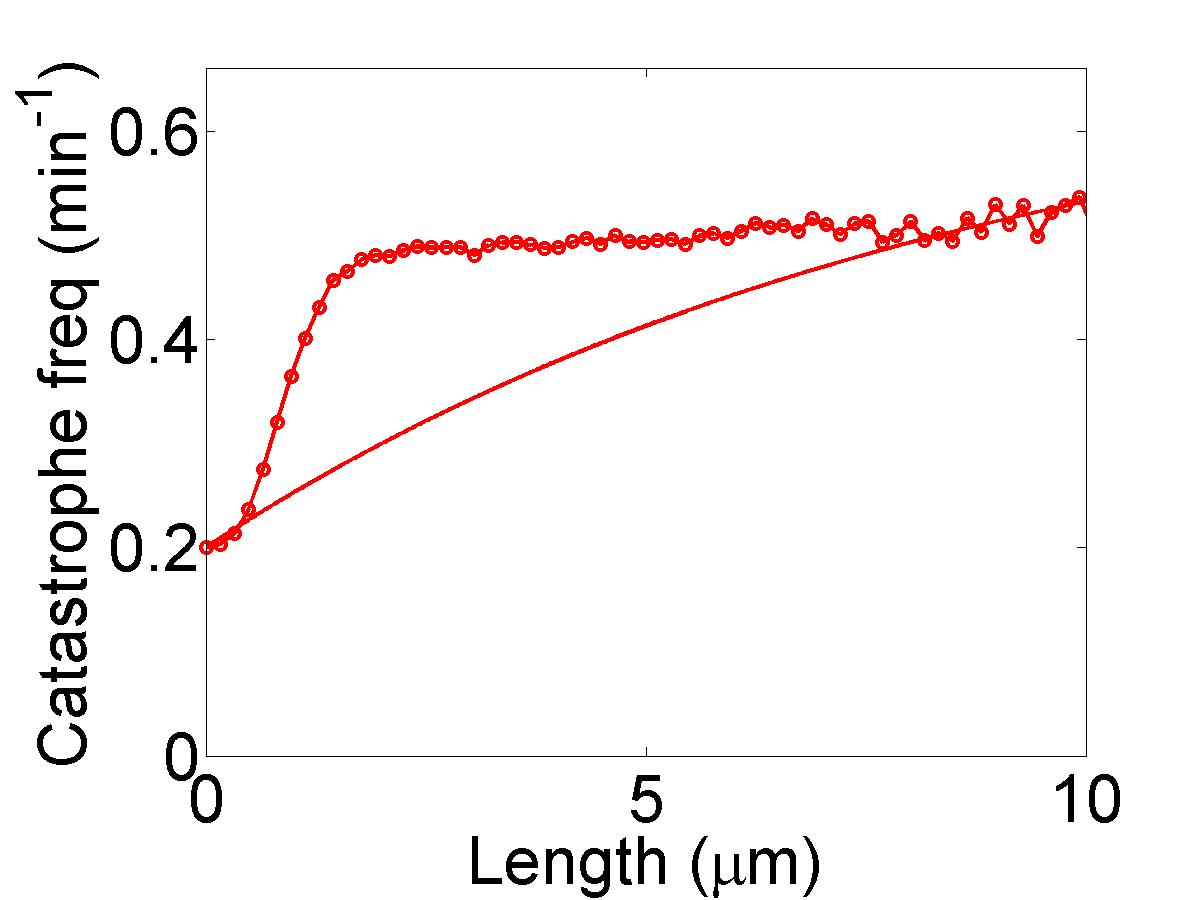}
 \includegraphics[width=5 cm]{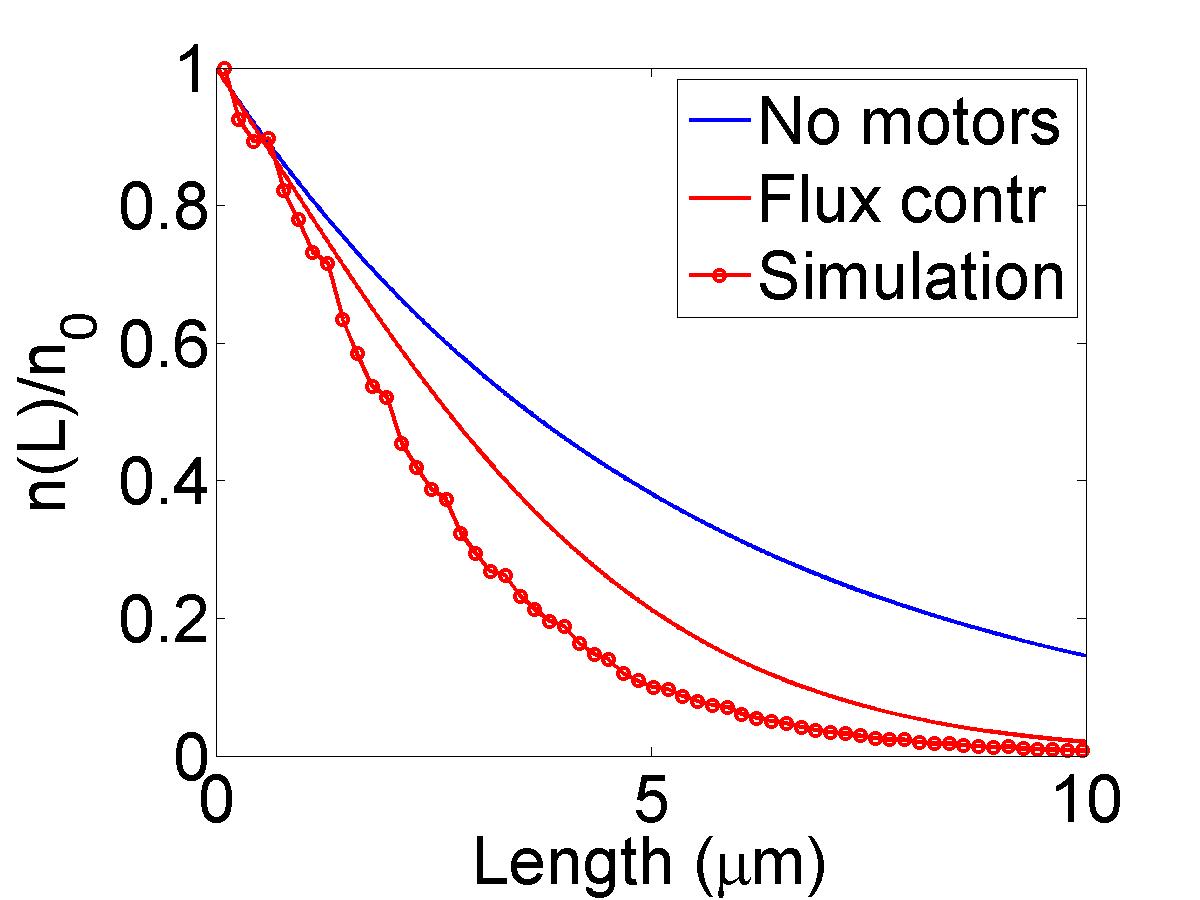}
 \caption{Dynamics of length regulation by altering catastrophe. Top,
   density-controlled model. Bottom, flux-controlled model. Left,
   example trace of filament length versus time in the stochastic
   simulation. Middle, catastrophe frequency as a function of filament
   length, comparing mean-field theory (line) and stochastic
   simulation (points).  Right, length distribution of dynamic
   filaments for the model with and without motors, comparing
   mean-field theory and stochastic simulation.  The presence of the
   motors leads to a significant decrease in the mean filament length.
   This figure uses the parameters $v = 3$ \mum \permin, $u = 1$ \mum
   \permin, $w = 7$ \mum \permin, minimum catastrophe frequency $f_c =
   0.2$ \permin, rescue frequency $f_r = 0.05$ \permin, $a = 8$ nm,
   and $\rhom = 125$ \permum. For the density-controlled model the
   parameters are $\kon = 1$ \pernm \permum \permin, $\koff = 0.25$
   \permin, $\koffe = 1.5$ \permin, bulk motor concentration $c = 2$
   nM, and $\alpha = 0.35$ \permin. The simulation of the
   density-controlled model has the same parameters except $\koffe =
   1$ \permin and $\alpha=0.38$ \permin.  For the flux-controlled
   model the parameters are $\kon = 3$ \pernm \permum \permin, $\koff
   = 0.25$ \permin, bulk motor concentration $c = 4$ nM, and $\alpha =
   7 \times 10^{-3}$. The simulation of the flux-controlled model has
   the same parameters as the corresponding mean-field theory except
   $\kon = 1.5$ \pernm \permum \permin\ and $\alpha = 2 \times
   10^{-3}$.}
  \label{fig:lengthdist}
\end{figure}

\subsection{Density-controlled catastrophe}

As above, we assume that the motors move faster than the filament
growth, so the motors track the end and the motor density is at steady
state.  In the density-controlled catastrophe model, we assume that
the catastrophe frequency increases linearly with the motor occupancy
at the end of the filament:
\begin{equation}
  \label{eq:catden}
  f = f_c + \alpha p_e.
\end{equation}
The motor occupancy at the end is determined by equation
\eqref{eq:dndt}.  At steady state, the occupancy at the end of the
growing filament is
\begin{equation}
  \label{eq:rhoec}
  p_e = \frac{b (1-e^{-L/\lambda})}{\koffe + b(1-e^{-L/\lambda})},
\end{equation}
with $b=(v-u)p_0/a$.  Using the integral
\begin{equation}
  \label{eq:lengthint}
  \int_0^L dL'  \frac{(1-e^{-L'/\lambda})}{\koffe +
    b(1-e^{-L'/\lambda})} =  \frac{L}{\koffe+b} - \frac{\lambda
    \koffe}{b(\koffe+b)} \ln \left(  1 +
    \frac{b}{\koffe}(1-e^{-L/\lambda})\right) , 
\end{equation}
the length distribution  is
\begin{equation}
  n(L) = n_{0} e^{-((f_c +\Delta f)/u-f_r/w)L}  \left( 
    1 +  \frac{(v-u)p_0}{a\koffe}
    (1-e^{-L/\lambda})\right)^{\frac{\alpha \lambda  
      \koffe}{ u (\koffe+(v-u) p_0/a)}} ,
\label{eq:densdist}
\end{equation}
where $\Delta f_{\rm den} = \alpha (v-u) p_0/(a \koffe+(v-u) p_0)$ is
the maximum possible increase in the catastrophe frequency in the
density-controlled model. We see two effects due to motors: first,
there is an effective increase in the catastrophe rate of $\Delta
f_{\rm den}$. Second, there is an additional multiplicative factor in
the length distribution. This factor approaches one in the limit of
short filaments ($L \ll \lambda$) and for typical experimental
parameters varies slowly with $L$.

Note that in this model the unbinding of the motor from end of the
filament is controlled by the rate constant $\koffe$.  This means that
a motor can processively track the end of a depolymerizing filament,
and this processivity tends to increase motor concentration at the end
of the filament and therefore enhance the filament-shortening effects
of motors.

\subsection{Flux-controlled catastrophe}

In the flux-controlled catastrophe model, we assume that the
catastrophe frequency increases linearly with the flux of motors to
the end:
\begin{equation}
  f = f_c + \alpha J.
\end{equation}
The flux to the end of the microtubule is $ J = p(L) \rhom (v-u)$.
Note that in this model $\alpha$ is dimensionless and that steric
interactions between motors that decrease the flux are neglected.  The
length distribution is then
\begin{equation}
  n = n_{0} e^{-(f_c/u + \Delta f /u-f_r/w)L} \exp\left(
    \frac{\alpha (v-u)\rhom p_0 \lambda}{u} (1-e^{-L/\lambda})
  \right) ,
\label{eq:fluxcont2}
\end{equation}
where $\Delta f_{\rm flux} = \alpha (v-u) \rhom p_0$ is the maximum
possible increase in the catastrophe frequency in the flux-controlled
model.  Again, we see two effects from the length-dependent
catastrophe: there is an effective increase in the catastrophe rate of
$\Delta f_{\rm flux}$, and there is an additional multiplicative
factor in the length distribution which is an exponential of an
exponential of the length.  This factor approaches 1 in the limit of
short microtubules ($L \ll \lambda$) and approaches the constant
factor $\exp\left(\frac{\alpha (v-u) \rhom p_0 \lambda}{u} \right)$ as
$L \to \infty$; for typical experimental parameters this factor is of
order 1.

We show simulations of filament length as a function of time,
calculations of the variation of catastrophe frequency with filament
length and filament length distributions in figure
\ref{fig:lengthdist}. We chose parameters from experiments on the
increase in catastrophe frequency associated with the kinesin-8 motor
Klp5/6 in fission yeast\cite{tischer09}, which found a catastrophe
frequency $f_c= 0.2$ \permin\ in cells lacking kinesin-8 motors and a
length-dependent increase in the catastrophe frequency up to a maximum
of $0.5$ \permin\ for filaments 8 \mum\ long in cells containing
motors.  With the correct choice of parameters, the length-dependent
increase in catastrophe frequency due to motors is qualitatively
similar to that measured by Tischer et al.~\cite{tischer09}.

The results of stochastic simulation are shown for comparison with
mean-field theory. For density-controlled catastrophe, there is
excellent agreement between stochastic simulation results and
mean-field theory if the parameters $\koffe$ and $\alpha$ are slightly
modified. The flux-controlled catastrophe model predictions of the
dependence of catastrophe frequency on filament length show only rough
qualitative agreement with mean-field theory; the shapes of the curves
are quite different. Even this level of agreement requires
modification of the parameters $\kon$ and $\alpha$.

\subsection{Mean filament length}

\begin{figure}[t]
  \centering
  \includegraphics[width=5 cm]{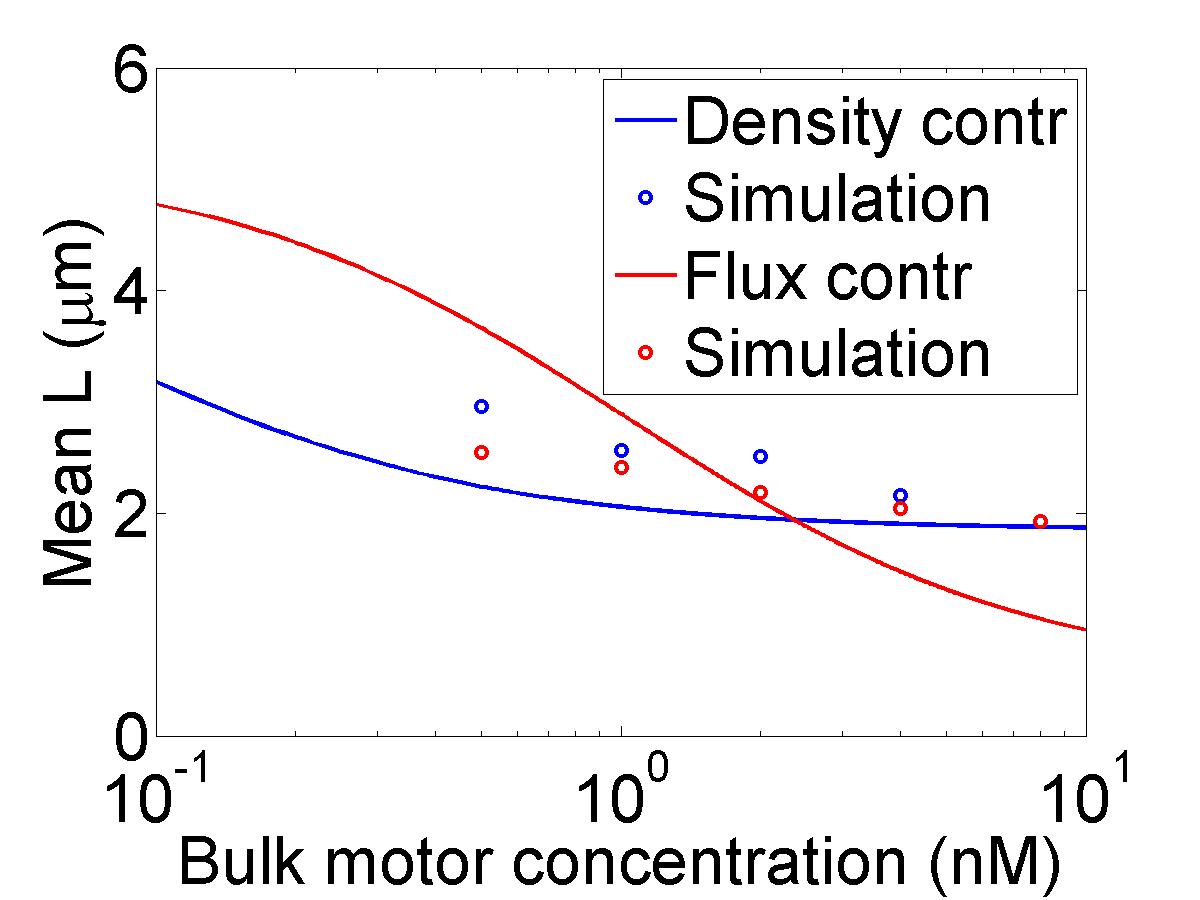}
  \includegraphics[width=5 cm]{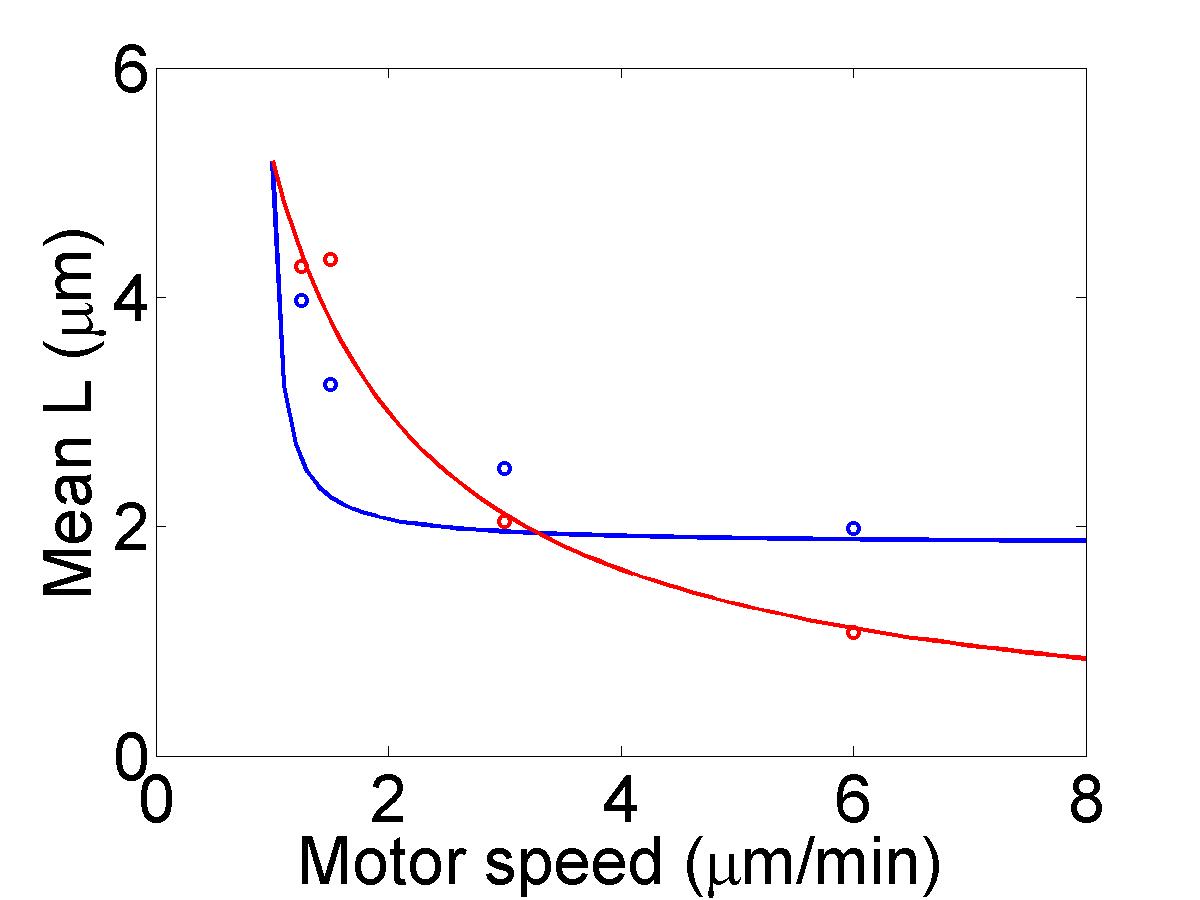}
  \includegraphics[width=5 cm]{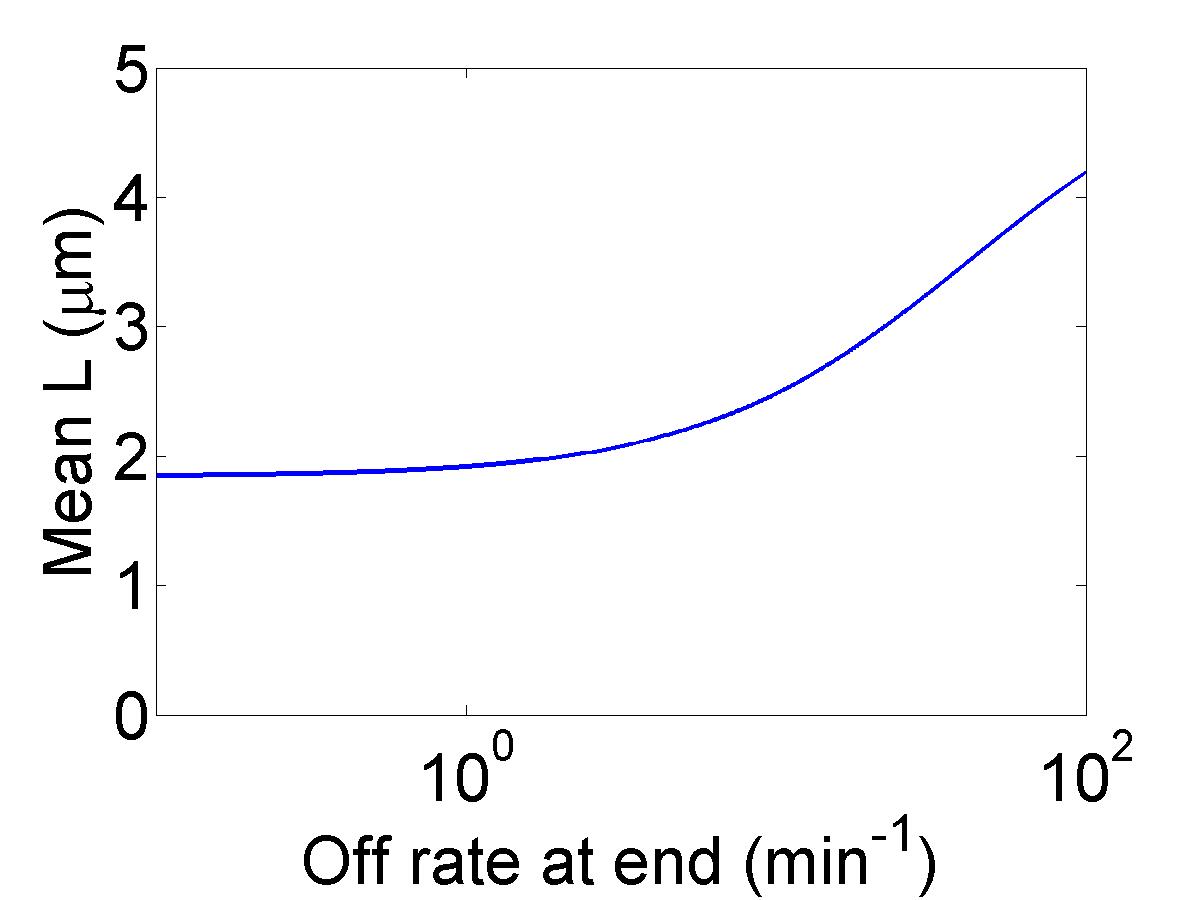}
  \includegraphics[width=5 cm]{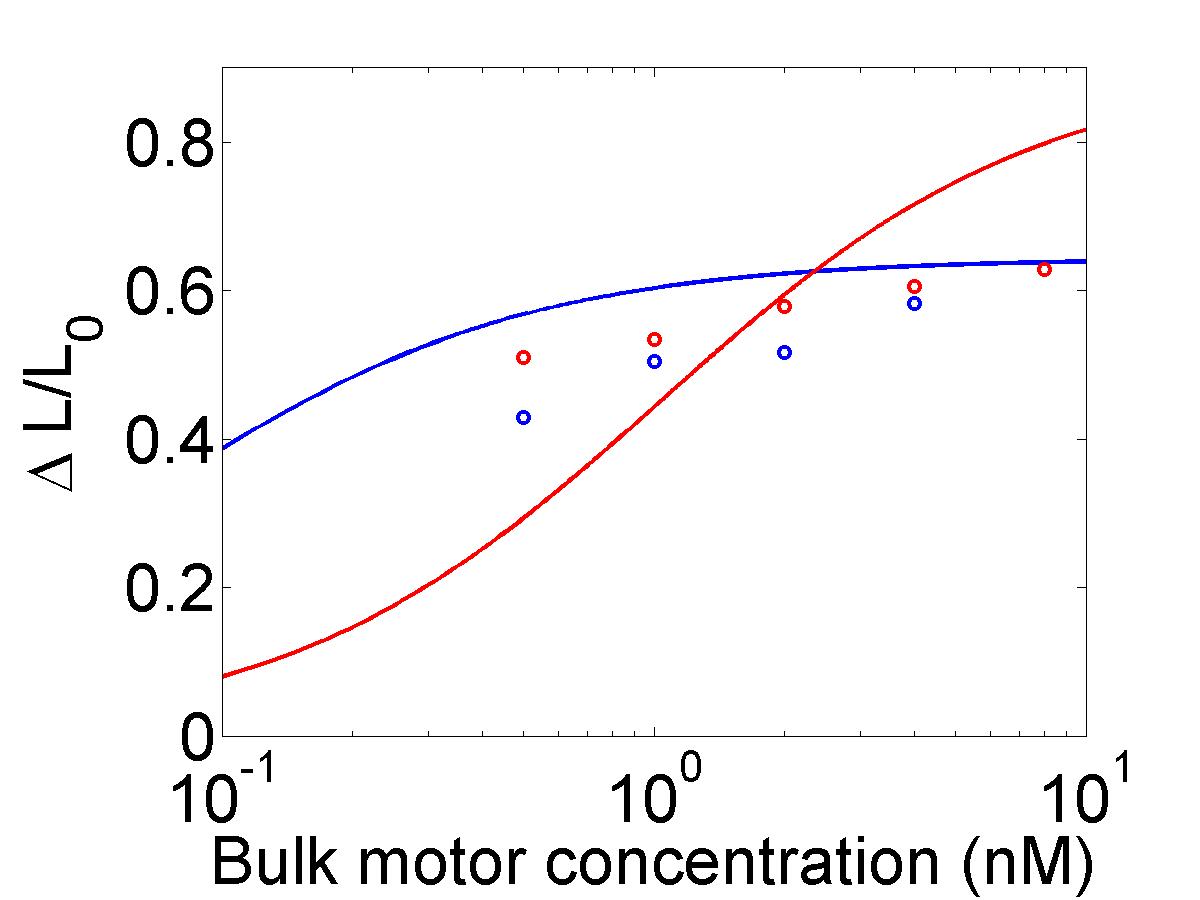}
  \includegraphics[width=5 cm]{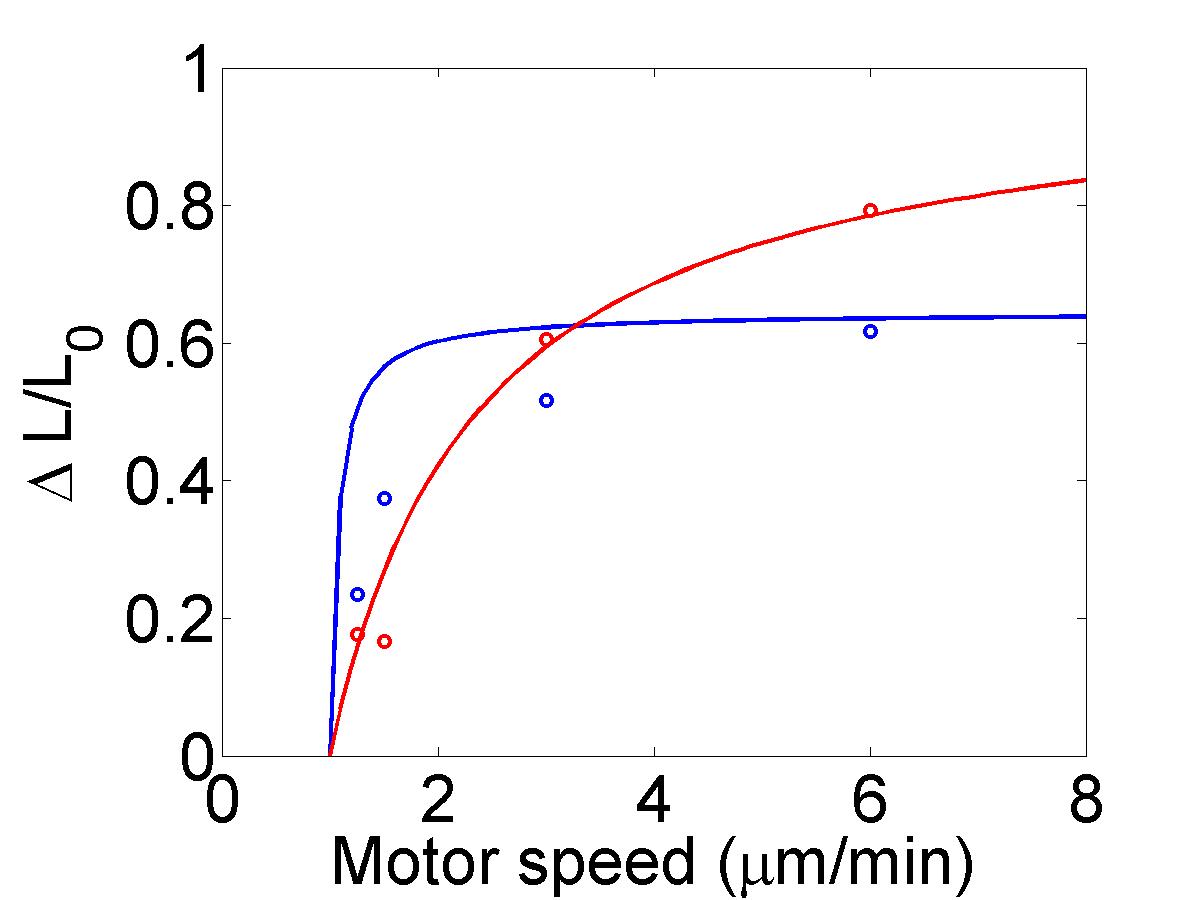}
  \includegraphics[width=5 cm]{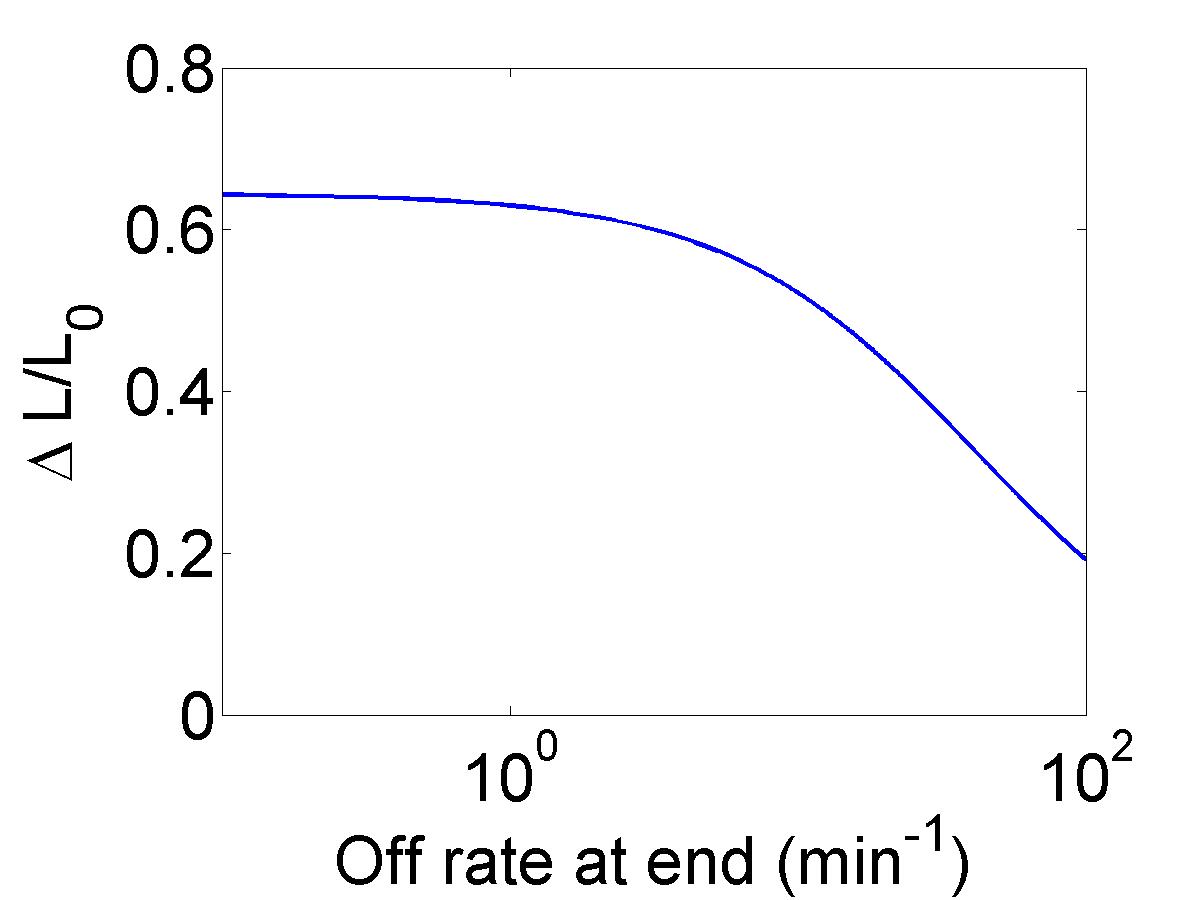}
  \caption{Mean filament length and changes in filament length as a
    function of motor parameters.  Left, variation as a function of
    bulk motor concentration.  Center, variation as a function of
    motor speed.  Right, variation as a function of $\koffe$ in the
    density-controlled model. This figure uses the same parameters as
    figure \ref{fig:lengthdist} except where noted in varying the bulk
    motor concentration and motor speed: $v = 3$ \mum \permin, $u = 1$
    \mum \permin, $w = 7$ \mum \permin, minimum catastrophe frequency
    $f_c = 0.2$ \permin, rescue frequency $f_r = 0.05$ \permin, $a =
    8$ nm, and $\rhom = 125$ \permum. For the density-controlled model
    the parameters are $\kon = 1$ \pernm \permum \permin, $\koff =
    0.25$ \permin, $\koffe = 1.5$ \permin, bulk motor
    concentration $c = 2$ nM, and $\alpha = 0.35$ \permin. The
    simulation of the density-controlled model has the same parameters
    except $\koffe = 1$ \permin\ and $\alpha=0.38$ \permin.  For the
    flux-controlled mean-field model the parameters are the same as
    for the density-controlled mean-field model except $\kon = 3$
    \pernm \permum \permin\ and $\alpha = 7 \times 10^{-3}$.  The
    simulation of the flux-controlled model has the same parameters as
    the corresponding mean-field theory except $\kon = 1.5$ \pernm
    \permum \permin\ and $\alpha = 2 \times 10^{-3}$.  }
  \label{fig:meanlength}
\end{figure}

The length-dependent catastrophe induced by motors changes the
microtubule length distribution in two ways: the effective catastrophe
frequency increases, and the length distribution is multiplied by an
additional function.  When the additional change in the functional
form due to this multiplication can be neglected, including only the
effective increase in the catastrophe frequency gives a simple result
for the change in mean filament length. The approximate length
distribution is
\begin{equation}
  \label{eq:ldistapprox2}
n(L) = n_0 e^{-[(f_c + \Delta f)/u-f_r/w] L }.
\end{equation}
We define $\bar{L}_0 = u w/(f_c w - f_r u)$, the mean filament length
in the absence of motors, and the mean length including motor effects
is $\bar{L} = \bar{L}_0-\Delta L$. The mean length is
\begin{equation}
  \label{eq:meanlength}
  \bar{L} = \bar{L}_0 \left(\frac{f_c - \frac{u}{w} f_r}{f_c -
      \frac{u}{w} f_r + \Delta f}  \right) 
\end{equation}
The fractional change in the mean length is 
\begin{equation}
  \label{eq:lengthchangerescue}
  \frac{\Delta L}{\bar{L}_0} = \frac{\Delta f}{f_c - \frac{u}{w} f_r + \Delta f}  
\end{equation}
This expression is an approximation that typically overestimates the
change in filament length due to motors, but has the advantage that it
has a simple analytical form that can be used to understand which
parameters control the mean length. The mean filament length is
related to the maximum increase in catastrophe frequency that can be
achieved by the motors. In the density-controlled model $\Delta f_{\rm
  den} = \alpha (v-u) p_0/(a \koffe+(v-u) p_0)$, and in the
flux-controlled model $\Delta f_{\rm flux} = \alpha (v-u) \rhom p_0$.

The change in the mean filament length varies with the bulk motor
concentration (through the occupancy $p_0$), the difference between
the filament growth speed and the motor walking speed, and the motor
unbinding rate from the filament end in the density-controlled model
(figure \ref{fig:meanlength}).  This
suggests that in cells length regulation could be tuned by altering
motor concentration, motor/filament velocity, or motor off rate at the
end of the filament in the density-controlled model. For typical
experimental parameters, the fractional change in mean filament length
varies from 0.1 to 0.9 with changes in these parameters.
While changes in bulk motor concentration can alter the mean filament
length, relatively large changes in the concentration (over two orders
of magnitude) lead to only modest changes in mean filament length. We
find that in comparison to stochastic simulation the approximate
expressions calculated above tend to overstate the changes in mean
filament length achievable through bulk motor density changes.
Particularly for the flux-controlled model, mean filament length is
more sensitive to alterations in motor speed. Varying motor speed over
a factor of 4 allows approximately factor of 10 change in the mean
filament length. Comparison to stochastic simulations shows that mean
filament length changes due to variation in motor speed are accurately
captured by the approximate model.

In the density-controlled model, the rate constant $\koffe$ controls
the residence time of a motor at the end of a filament. In the limit
that this rate becomes large compared to other rates in the problem,
the motors rapidly unbind upon reaching the filament end and the
changes in filament length due to motors become smaller.

The increase in catastrophe due to motors can shift the filament
length distribution from the unbounded growth regime to the bounded
growth regime. If $f_r/w > f_c/u$, the filament dynamics are in the
unbounded growth regime, with no defined mean filament length. If the
increase in catastrophe due to motors $\Delta f$ is large enough, the
presence of motors can shift the distribution back to the bounded
growth regime characterized by an exponential length distribution and
a well-defined mean filament length. This requires that $\Delta f >
f_r u/w-f_c$.  This relationship implies a minimum motor concentration
to shift from  unbounded to bounded growth.

\section{Stochastic simulation}
\label{sim}

We developed a kinetic Monte Carlo simulation of the model as shown in
figure \ref{fig:model}. The model considers a single filament
(equivalent to representing a microtubule by a single protofilament)
made up of a varying number of monomers. Each motor occupies a single
filament monomer. Typical time scales of motor and filament processes
are of order seconds to minutes, so we chose a simulation time step of
0.01 s.  At each time step a number of monomers equal to the total
number of monomers currently in the the filament is randomly sampled
and one step of the polymerization dynamics at the end of the filament
is performed.  Each site on the filament except the last site has the
same rules: a motor can bind to an empty site, if the next site
forward is empty a motor can step forward, and a motor can unbind to
create an empty site.

The behavior of the last site (the end of the filament) varies
depending on the model considered. In the model of length regulation
by depolymerization (figure \ref{fig:model}A), the filament can grow
by addition of a monomer. Growth is independent of motor occupancy at
the last site. The filament can shrink by one monomer depending on the
motor occupancy at the end of the filament. In the
\textbf{density-controlled depolymerization} model, removal of the
terminal monomer can occur if the terminal site is occupied by a
motor. In this model the motor can processively track the
depolymerizing filament: if the penultimate site is empty, the motor
at the last site steps backward when the terminal monomer is removed.
If the second-to-last site is occupied, the motor on the terminal site
is removed when the terminal monomer is removed. The motor at the
terminal site can also directly unbind from the filament without
removal of a monomer. In the \textbf{flux-controlled depolymerization}
model, removal of the terminal monomer occurs when the last and
penultimate sites are both occupied by motors and the motor at the
penultimate site attempts a forward step. In this case the terminal
motor and monomer are both removed.

In the model of length regulation by catastrophe (figure
\ref{fig:model}B), the filament stochastically switches between
growing and shrinking states. Speeds of growth and shrinkage as well
as the rescue frequency are independent of motor occupancy at the last
site. The catastrophe frequency is increased depending on the motor
occupancy at the end of the filament. In the
\textbf{density-controlled catastrophe} model, the catastrophe
frequency is increased by $\alpha$ if the terminal site is occupied by
a motor. In this model the motor can processively track the
depolymerizing filament: if the filament is shortening and the
penultimate site is empty, the motor at the last site steps backward
when the terminal monomer is removed. If the second-to-last site is
occupied, the motor on the terminal site is removed when the terminal
monomer is removed.  The motor at the terminal site can also directly
unbind from the filament without removal of a monomer. In the
\textbf{flux-controlled catastrophe} model, the increase in the
catastrophe frequency by $\alpha$ occurs when the last and penultimate
sites are both occupied by motors and the motor at the penultimate
site attempts a forward step. In this case the terminal motor is
removed independent of whether or not a catastrophe occurs.

In all versions of the model, if the filament fully depolymerizes (0
sites remain) a new filament is nucleated containing 1 site.  For each
parameter set we performed 5-10 simulations of $10^6-10^8$ time steps.

\section{Conclusion}

We have considered an example of biophysical length regulation by
motors that walk along a filament and promote filament shortening,
inspired by experiments on the effects of kinesin-8 motor proteins on
microtubule dynamics. The motors bind to microtubules and move toward
their plus ends, and the presence of motors at the filament end alters
microtubule polymerization dynamics.

The first mechanism we considered is a simplified model of length
regulation, in which the motors directly catalyze depolymerization of
the filament from its plus end. When the action of the motors is
balanced by a constant filament polymerization rate, a steady-state
filament length can be reached. This mechanism neglects any
fluctuating filament dynamics: only a single steady-state length is
reached. In the stochastic simulation, fluctuations due to stochastic
motor/filament dynamics lead to a spread about the mean length but the
distribution of filament lengths remains strongly peaked.  There is a
minimum bulk motor concentration on the filament to reach a
steady-state length, because if there are too few motors,
motor-induced depolymerization can never be fast enough to balance the
intrinsic polymerization. In addition, inequalities involving the
motor motion constrain the parameter regime where steady-state
solutions are possible. The steady-state filament length differs from
the length scale $\lambda$ which characterizes the motor density
profile. The steady-state filament length depends sensitively on the
bulk motor concentration, implying that this mechanism of length
regulation requires tight control of total motor number to operate
successfully.

Other recent theory work has addressed length regulation due to
depolymerizing motors and filament kinetics described by constant
growth \cite{govindan08,melbinger12} or treadmilling \cite{johann12}.
Govindan et al.~considered a similar model for motor motion, but used
an absorbing boundary condition for motors at the filament plus end, an
approximation that doesn't apply to filaments with biologically
realistic growth and shrinkage rates. Their work found an exponential
filament length distribution for typical parameter values corresponding to
Kip3 \cite{govindan08}.  Melbinger et al.~improved the model of
Govindan et al.~by studying in detail how effects of motor crowding
near the microtubule end control the depolymerization dynamics. They
discovered a parameter regime in which filament length is well regulated,
and how the length depends on motor kinetics \cite{melbinger12}.
Johann et al.~considered the related problem of how length regulation
can be achieved by depolymerizing motors on filaments that undergo
treadmilling dynamics (addition of subunits at one end and removal at
the other) \cite{johann12}.

In the model of length regulation by depolymerization we have
discussed, constant growth is balanced by length-dependent
depolymerization.  In the balance-point model of flagellar length
regulation in \textit{Chlamydomonas}, a constant rate of flagellar
disassembly is balanced by a length-dependent rate of flagellar
assembly, leading to a fixed flagellar length
\cite{engel09,marshall05}. While the underlying microscopic mechanisms
of flagellar length regulation differ from those discussed here, the
conceptual similarity is striking. Perhaps this basic idea of
regulating length by making assembly or disassembly length-dependent
while the other process (disassembly or assembly) is length
independent could be a general paradigm for length regulation, at
least of microtubule-based structures.

A more biologically relevant model for the length regulation of
dynamic microtubules is length regulation by altering catastrophe, in
which the filament undergoes dynamic instability characterized by
long-lived growing and shrinking states with transitions between
growth and shrinkage. The effect of the motors at the end is then not
to directly depolymerize the filament but to increase the catastrophe
frequency. We calculate how the filament length distribution is
altered by the motor-dependent increase in catastrophe frequency, and
derive a simple approximate expression that relates the mean filament
length to the maximum increase in catastrophe frequency that can be
achieved by the motors. The mean filament length varies modestly with
bulk motor concentration but is sensitive to the difference between
the filament growth speed and the motor walking speed.

The increase in catastrophe frequency associated with the kinesin-8
motor Klp5/6 in fission yeast cells was measured by Tischer et
al.~\cite{tischer09}, who found a catastrophe frequency $f_c= 0.2$
\permin\ in cells lacking Klp5/6 and a length-dependent increase in
the catastrophe frequency up to a maximum of $0.5$ \permin\ for
filaments 8 \mum\ long in cells containing motors.  With the correct
choice of parameters, our model displays a length-dependent increase
in catastrophe frequency due to motors which is qualitatively similar
to that measured by Tischer et al.  Using these parameters in our
model, changes in the mean length of a factor of 2 can be achieved by
this mechanism.

\begin{acknowledgements} The authors thank Robert Blackwell, Matt Glaser, Loren Hough and
Dick McIntosh for useful discussions. This work was supported by NSF
CAREER Award DMR-0847685, NSF MRSEC Grant DMR-0820579, and NIH
training grant T32 GM-065103.
\end{acknowledgements}

\section*{References}
\bibliography{kinesin8,zoterolibrary}{}

\begin{thebibliography}{10}

\bibitem{day00}
S.~J. Day and P.~A. Lawrence.
\newblock Measuring dimensions: the regulation of size and shape.
\newblock {\em Development}, 127(14):2977{\textendash}2987, 2000.

\bibitem{hafen03}
E.~Hafen and H.~Stocker.
\newblock How are the sizes of cells, organs, and bodies controlled?
\newblock {\em {PLoS} Biology}, 1(3):e86, 2003.

\bibitem{cook07}
Mike Cook and Mike Tyers.
\newblock Size control goes global.
\newblock {\em Current Opinion in Biotechnology}, 18(4):341--350, August 2007.

\bibitem{goshima05}
G.~Goshima, R.~Wollman, N.~Stuurman, J.~M. Scholey, and R.~D. Vale.
\newblock Length control of the metaphase spindle.
\newblock {\em Current Biology}, 15(22):1979{\textendash}1988, 2005.

\bibitem{walczak96}
Claire~E. Walczak, Timothy~J. Mitchison, and Arshad Desai.
\newblock {XKCM1:} a xenopus kinesin-related protein that regulates microtubule
  dynamics during mitotic spindle assembly.
\newblock {\em Cell}, 84(1):37--47, January 1996.

\bibitem{rivero96}
F.~Rivero, B.~Koppel, B.~Peracino, S.~Bozzaro, F.~Siegert, C.~J. Weijer,
  M.~Schleicher, R.~Albrecht, and A.~A. Noegel.
\newblock The role of the cortical cytoskeleton: F-actin crosslinking proteins
  protect against osmotic stress, ensure cell size, cell shape and motility,
  and contribute to phagocytosis and development.
\newblock {\em Journal of Cell Science}, 109(11):2679--2691, November 1996.

\bibitem{revenu04}
C.~Revenu, R.~Athman, S.~Robine, and D.~Louvard.
\newblock The co-workers of actin filaments: from cell structures to signals.
\newblock {\em Nature Reviews Molecular Cell Biology},
  5(8):635{\textendash}646, 2004.

\bibitem{dogterom93}
Marileen Dogterom and Stanislas Leibler.
\newblock Physical aspects of the growth and regulation of microtubule
  structures.
\newblock {\em Physical Review Letters}, 70(9):1347--1350, March 1993.

\bibitem{akhmanova05}
Anna Akhmanova and Casper~C Hoogenraad.
\newblock Microtubule plus-end-tracking proteins: mechanisms and functions.
\newblock {\em Current Opinion in Cell Biology}, 17(1):47--54, February 2005.

\bibitem{marshall05}
Wallace~F. Marshall, Hongmin Qin, M\'{o}nica~Rodrigo Brenni, and Joel~L.
  Rosenbaum.
\newblock Flagellar length control system: Testing a simple model based on
  intraflagellar transport and turnover.
\newblock {\em Molecular Biology of the Cell}, 16(1):270--278, January 2005.

\bibitem{gupta06}
M.~L. Gupta, P.~Carvalho, D.~M. Roof, and D.~Pellman.
\newblock Plus end-specific depolymerase activity of {K}ip3, a kinesin-8
  protein, explains its role in positioning the yeast mitotic spindle.
\newblock {\em Nature Cell Biology}, 8(9):913--923, 2006.

\bibitem{varga06}
V.~Varga, J.~Helenius, K.~Tanaka, A.~A. Hyman, T.~U. Tanaka, and J.~Howard.
\newblock Yeast kinesin-8 depolymerizes microtubules in a length-dependent
  manner.
\newblock {\em Nature Cell Biology}, 8(9):957--962, 2006.

\bibitem{varga09}
Vladimir Varga, Cecile Leduc, Volker Bormuth, Stefan Diez, and Jonathon Howard.
\newblock {Kinesin-8 Motors Act Cooperatively to Mediate Length-Dependent
  Microtubule Depolymerization}.
\newblock {\em Cell}, 138(6):1174, 2009.

\bibitem{hough09}
L.~E. Hough, A.~Schwabe, M.~A. Glaser, J.~R. {McIntosh}, and M.~D. Betterton.
\newblock Microtubule depolymerization by the kinesin-8 motor kip3p: a
  mathematical model.
\newblock {\em Biophysical Journal}, 96(8):3050{\textendash}3064, 2009.

\bibitem{west01}
Robert~R West, Terra Malmstrom, Cynthia~L Troxell, and J~R McIntosh.
\newblock {Two related kinesins, klp5+ and klp6+, foster microtubule
  disassembly and are required for meiosis in fission yeast.}
\newblock {\em Molecular Biology of the Cell}, 12(12):3919--32, 2001.

\bibitem{west02}
Robert~R West, Terra Malmstrom, and J~Richard McIntosh.
\newblock {Kinesins klp5(+) and klp6(+) are required for normal chromosome
  movement in mitosis.}
\newblock {\em Journal of Cell Science}, 115:931--40, 2002.

\bibitem{garcia02a}
Miguel~Angel Garcia, Nirada Koonrugsa, and Takashi Toda.
\newblock {Two kinesin-like Kin I family proteins in fission yeast regulate the
  establishment of metaphase and the onset of anaphase A.}
\newblock {\em Current biology}, 12(8):610--21, 2002.

\bibitem{garcia02b}
M.~A. Garcia, N.~Koonrugsa, and T.~Toda.
\newblock Spindle-kinetochore attachment requires the combined action of {K}in
  {I}-like {K}lp5/6 and {A}lp14/{D}is1-{MAP}s in fission yeast.
\newblock {\em EMBO Journal}, 21(22):6015--6024, 2002.

\bibitem{savoian04}
M.~S. Savoian, M.~K. Gatt, M.~G. Riparbelli, G.~Callaini, and D.~M. Glover.
\newblock Drosophila {Klp67A} is required for proper chromosome congression and
  segregation during meiosis i.
\newblock {\em Journal of Cell Science}, 117(16):3669{\textendash}3677, 2004.

\bibitem{mayr07}
M.~I. Mayr, S.~Hummer, J.~Bormann, T.~Gruner, S.~Adio, G.~Woehlke, and T.~U.
  Mayer.
\newblock The human kinesin {K}if18{A} is a motile microtubule depolymerase
  essential for chromosome congression.
\newblock {\em Current Biology}, 17(6):488--498, 2007.

\bibitem{jaqaman10}
K.~Jaqaman, E.~M. King, A.~C. Amaro, J.~R. Winter, J.~F. Dorn, H.~L. Elliott,
  N.~Mchedlishvili, S.~E. {McClelland}, I.~M. Porter, M.~Posch, et~al.
\newblock Kinetochore alignment within the metaphase plate is regulated by
  centromere stiffness and microtubule depolymerases.
\newblock {\em The Journal of Cell Biology}, 188(5):665{\textendash}679, 2010.

\bibitem{gandhi04}
R.~Gandhi, S.~Bonaccorsi, D.~Wentworth, S.~Doxsey, M.~Gatti, and A.~Pereira.
\newblock The drosophila kinesin-like protein {KLP67A} is essential for mitotic
  and male meiotic spindle assembly.
\newblock {\em Molecular Biology of the Cell}, 15(1):121{\textendash}131, 2004.

\bibitem{gatt05}
M.~K. Gatt, M.~S. Savoian, M.~G. Riparbelli, C.~Massarelli, G.~Callaini, and
  D.~M. Glover.
\newblock {Klp67A} destabilises pre-anaphase microtubules but subsequently is
  required to stabilise the central spindle.
\newblock {\em Journal of Cell Science}, 118(12):2671{\textendash}2682, 2005.

\bibitem{unsworth08}
Amy Unsworth, Hirohisa Masuda, Susheela Dhut, and Takashi Toda.
\newblock {Fission yeast kinesin-8 Klp5 and Klp6 are interdependent for mitotic
  nuclear retention and required for proper microtubule dynamics}.
\newblock {\em Molecular Biology of the Cell}, 19(12):5104, 2008.

\bibitem{tischer09}
Christian Tischer, Damian Brunner, and Marileen Dogterom.
\newblock Force- and kinesin-8-dependent effects in the spatial regulation of
  fission yeast microtubule dynamics.
\newblock {\em Molecular Systems Biology}, 5, March 2009.

\bibitem{du10}
Y.~Du, C.~A. English, and R.~Ohi.
\newblock The kinesin-8 {Kif18A} dampens microtubule plus-end dynamics.
\newblock {\em Current Biology}, 20(4):374{\textendash}380, 2010.

\bibitem{peters10}
Carsten Peters, Katju{\textbar}[scaron]{\textbar}a Brejc, Lisa Belmont,
  Andrew~J. Bodey, Yan Lee, Ming Yu, Jun Guo, Roman Sakowicz, James Hartman,
  and Carolyn~A. Moores.
\newblock Insight into the molecular mechanism of the multitasking kinesin-8
  motor.
\newblock {\em The {EMBO} Journal}, 29(20):3437--3447, September 2010.

\bibitem{wang10}
H.~Wang, I.~Brust-Mascher, D.~Cheerambathur, and J.M. Scholey.
\newblock {Coupling between microtubule sliding, plus-end growth and spindle
  length revealed by kinesin-8 depletion}.
\newblock {\em Cytoskeleton}, 67(11):715--728, 2010.

\bibitem{gardner11a}
Melissa~K. Gardner, Marija Zanic, Christopher Gell, Volker Bormuth, and
  Jonathon Howard.
\newblock Depolymerizing kinesins {K}ip3 and {MCAK} shape cellular microtubule
  architecture by differential control of catastrophe.
\newblock {\em Cell}, 147(5):1092--1103, November 2011.

\bibitem{masuda11}
Natsuko Masuda, Tetsuhiro Shimodaira, Shu-Jen Shiu, Noriko Tokai-Nishizumi,
  Tadashi Yamamoto, and Miho Ohsugi.
\newblock {Microtubule stabilization triggers the plus-end accumulation of
  Kif18A/kinesin-8.}
\newblock {\em Cell Structure and Function}, 36(2):261--7, January 2011.

\bibitem{mayr11}
Monika~I. Mayr, Marko Storch, Jonathon Howard, and Thomas~U. Mayer.
\newblock A non-motor microtubule binding site is essential for the high
  processivity and mitotic function of kinesin-8 {Kif18A}.
\newblock {\em {PLoS} One}, 6(11):e27471, November 2011.

\bibitem{stumpff11}
Jason Stumpff, Yaqing Du, Chauca~A. English, Zoltan Maliga, Michael Wagenbach,
  Charles~L Asbury, Linda Wordeman, and Ryoma Ohi.
\newblock A tethering mechanism controls the processivity and
  kinetochore-microtubule plus-end enrichment of the kinesin-8 {K}if18{A}.
\newblock {\em Molecular cell}, 43(5):764--75, September 2011.

\bibitem{su11}
X.~Su, W.~Qiu, M.L. {Gupta Jr}, J.B. Pereira-Leal, S.L. Reck-Peterson, and
  D.~Pellman.
\newblock {Mechanisms underlying the dual-mode regulation of microtubule
  dynamics by {K}ip3/kinesin-8.}
\newblock {\em Molecular cell}, 43(5):751, 2011.

\bibitem{weaver11}
L.N. Weaver, S.C. Ems-McClung, J.R. Stout, C.~LeBlanc, S.L. Shaw, M.K. Gardner,
  and C.E. Walczak.
\newblock {Kif18A} uses a microtubule binding site in the tail for plus-end
  localization and spindle length regulation.
\newblock {\em Current Biology}, 2011.

\bibitem{erent12}
M.~Erent, D.~R. Drummond, and R.~A. Cross.
\newblock S. pombe kinesins-8 promote both nucleation and catastrophe of
  microtubules.
\newblock {\em {PloS} One}, 7(2):e30738, 2012.

\bibitem{grissom09}
P.~M. Grissom, T.~Fiedler, E.~L. Grishchuk, D.~Nicastro, R.~R. West, and J.~R.
  {McIntosh}.
\newblock Kinesin-8 from fission yeast: A heterodimeric,
  plus-end{\textendash}directed motor that can couple microtubule
  depolymerization to cargo movement.
\newblock {\em Molecular Biology of the Cell}, 20(3):963{\textendash}972, 2009.

\bibitem{brun09}
Ludovic Brun, Beat Rupp, Jonathan~J. Ward, and Fran\c{c}ois N\'{e}d\'{e}lec.
\newblock A theory of microtubule catastrophes and their regulation.
\newblock {\em Proceedings of the National Academy of Sciences},
  106(50):21173--21178, December 2009.

\bibitem{reese11}
Louis Reese, Anna Melbinger, and Erwin Frey.
\newblock Crowding of molecular motors determines microtubule depolymerization.
\newblock {\em Biophysical Journal}, 101(9):2190--2200, November 2011.

\bibitem{govindan08}
B.~S. Govindan, M.~Gopalakrishnan, and D.~Chowdhury.
\newblock Length control of microtubules by depolymerizing motor proteins.
\newblock {\em Europhysics Letters}, 83(4):40006, August 2008.

\bibitem{johann12}
Denis Johann, Christoph Erlenk\"{a}mper, and Karsten Kruse.
\newblock Length regulation of active biopolymers by molecular motors.
\newblock {\em Physical Review Letters}, 108(25):258103, June 2012.

\bibitem{melbinger12}
Anna Melbinger, Louis Reese, and Erwin Frey.
\newblock Microtubule length regulation by molecular motors.
\newblock {\em Physical Review Letters}, 108(25):258104, June 2012.

\bibitem{tischer10}
Christian Tischer, Pieter~Rein ten Wolde, and Marileen Dogterom.
\newblock Providing positional information with active transport on dynamic
  microtubules.
\newblock {\em Biophysical Journal}, 99(3):726--735, August 2010.

\bibitem{parmeggiani04}
A.~Parmeggiani, T.~Franosch, and E.~Frey.
\newblock Totally asymmetric simple exclusion process with {Langmuir} kinetics.
\newblock {\em Physical Review E}, 70(4), October 2004.

\bibitem{nowak07}
Sarah~A. Nowak, Pak-Wing Fok, and Tom Chou.
\newblock Dynamic boundaries in asymmetric exclusion processes.
\newblock {\em Physical Review E}, 76(3):031135--11, 2007.

\bibitem{engel09}
Benjamin~D. Engel, William~B. Ludington, and Wallace~F. Marshall.
\newblock Intraflagellar transport particle size scales inversely with
  flagellar length: revisiting the balance-point length control model.
\newblock {\em The Journal of Cell Biology}, 187(1):81--89, October 2009.

\end{thebibliography}
\bibliographystyle{unsrt}

\end{document}